\documentclass[useAMS,usenatbib]{mn2e}
\RequirePackage[loading]{tracefnt}
\usepackage{epsfig}
\usepackage{amssymb}
\usepackage{amsmath}
\usepackage{natbib}
\usepackage{threeparttable} 
\usepackage{graphicx}
\usepackage{setspace}
\usepackage{longtable}
\usepackage{multicol}
\usepackage{subfigure}
\usepackage{verbatim}
\usepackage{rotating}
\usepackage{lscape}
\newcommand\apj{ApJ}
\newcommand\apjl{ApJ}
\newcommand\apjs{ApJS}
\newcommand\aap{A\&A}
\newcommand\mnras{MNRAS}
\newcommand\nat{Nature}
\newcommand\ssr{Space Sci. Rev.}%
\def\hide#1{}

\title[X-ray bursts in 4U 1728--34]{The link between coherent burst 
oscillations, burst spectral evolution and accretion state in 4U 1728--34}

\author[G. Zhang et al.]{Guobao Zhang$^{1, 2}$\thanks{E-mail: guobao.zhang@nyu.edu}, Mariano M\'endez$^{2}$, Michael Zamfir$^{3}$, 
Andrew Cumming$^{3}$ \\
$^{1}$New York University Abu Dhabi, P.O. Box 129188, Abu Dhabi, United Arab Emirates\\
$^{2}$Kapteyn Astronomical Institute, University of Groningen, P.O. BOX
800, 9700 AV Groningen, The Netherlands\\ 
$^{3}$Department of Physics and McGill Space Institute, McGill University, 3600 rue University, Montreal, QC H3A 2T8, Canada
}

\begin{document}

\maketitle

\label{firstpage}

\date{Accepted. Received; in original form}
 
\begin{abstract}

Coherent oscillations and the evolution of the X-ray spectrum during thermonuclear
X-ray bursts in accreting neutron-star X-ray binaries have been studied intensively
but separately.  We analysed all the X-ray bursts of the source 4U 1728--34 with the
Rossi X-ray Timing Explorer. 
We found that the presence of burst oscillations can be used to predict the 
behaviour of the blackbody radius during the cooling phase of the bursts.  If a 
burst shows oscillations, during the cooling phase the blackbody 
radius remains more or less constant for $\sim 2 - \sim 8$ s, whereas in bursts 
that do not show oscillations the blackbody radius either remains constant for 
more than $\sim 2 - \sim 8$ s or it shows a rapid (faster than $\sim 2$ s) 
decrease and increase.
Both the presence of burst oscillations and the time-dependent spectral behaviour of
the bursts are affected by accretion rate. We also found that the rise time and
convexity of the bursts' light curve are different in bursts with and without
oscillations in 4U 1728--34. Bursts with oscillations have a short rise time 
($\sim 0.5$ s) and show both positive and negative convexity, whereas bursts 
without oscillations have a long rise time  ($\sim 1$ s) and mostly positive 
convexity. This is consistent with the idea that burst oscillations are associated 
with off-equator ignition.

\end{abstract}

\begin{keywords}
stars: neutron --- X-rays: binaries --- X-rays: bursts --- stars:
individual: 4U 1728--34
\end{keywords}

\section{introduction}
\label{introduction}

Thermonuclear, type-I, X-ray bursts \citep[e.g.,][]{Lewin93, Strohmayer03, Galloway08a} 
are due to unstable burning of accreted H and/or He on the surface of accreting 
neutron stars (NS) in low-mass X-ray binaries (LMXBs). Typical bursts exhibit rise 
times of 1$-$10 s, durations of a few tens of seconds to a few minutes, and have total 
energy outputs of $10^{39}-10^{40}$ erg. Some X-ray bursts are strong enough to
reach the Eddington luminosity.  In those so-called photospheric radius expansion (PRE) 
bursts, the radiation pressure is high enough to trigger the expansion of the outer 
layers of the NS atmosphere \citep[e.g.,][]{Basinska84, Kuulkers02}.

The inferred emission area of a NS can be estimated from the fitting of the energy 
spectra during the decaying phase of type-I X-ray bursts, assuming that the 
thermonuclear flash expands to cover the entire star during the radius expansion 
and cooling phases of the burst \citep{Paradijs78}. There are a number of  theoretical 
and observational arguments that support this assumption \citep[see, e.g. ][]{Fryxel82, 
Bildsten95, Spitkovsky02, Strohmayer03}. Recent work has shown that during the cooling phase 
of the bursts, the relation between the bolometric flux and the temperature is significantly 
different from the canonical $F\propto T^4$ relation that is expected if the apparent 
emitting area on the surface of the NS remains constant as the flux decreases during the 
decay of the burst and if the spectrum is blackbody \citep{Zhanggb11, Suleimanov11, Garcia13}. 
This could be due to either changes in the emitting area of the NS, or to changes in the 
colour-correction factor, $f_{\rm c}$, which accounts for hardening of the spectrum arising 
from electron scattering in the NS atmosphere, among other factors \citep{Paradijs82, Suleimanov11}.

\begin{figure*}
    \centering
        \includegraphics[width=7.0in,angle=0]{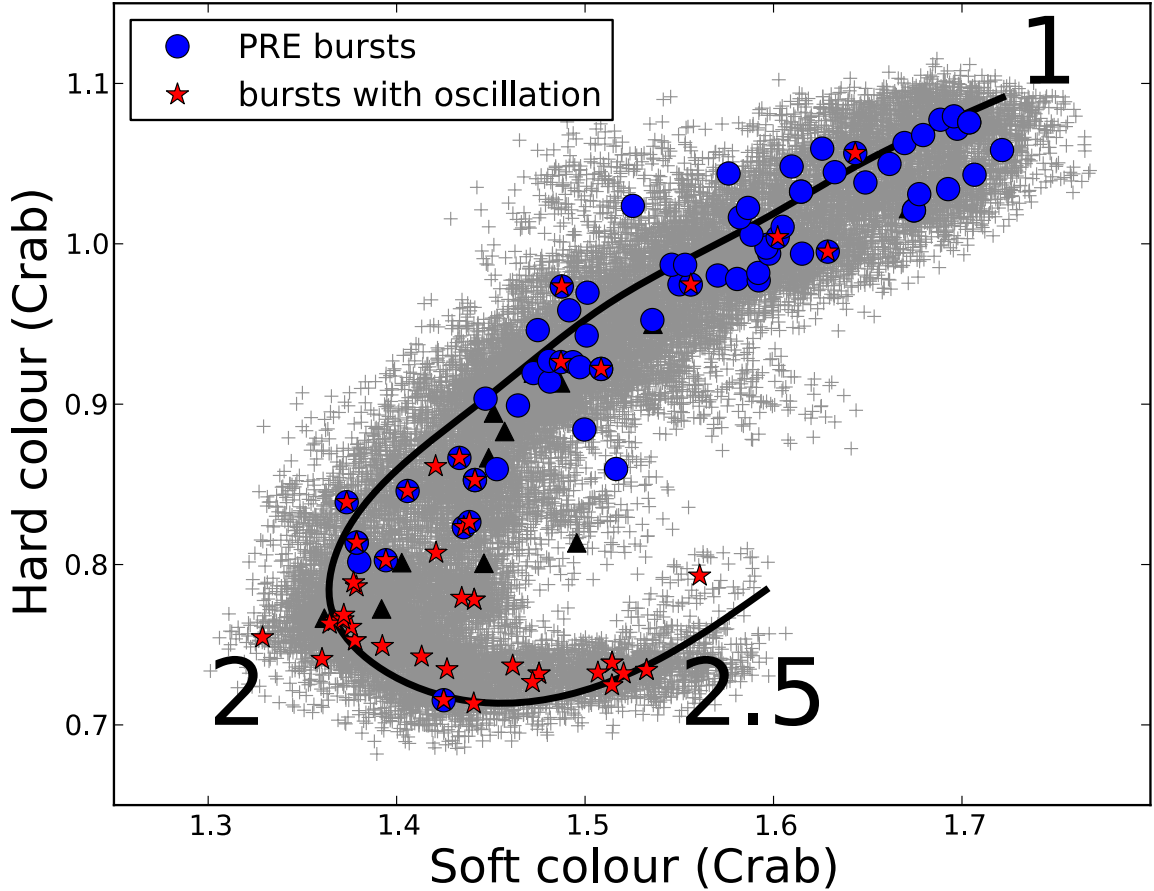}
    \caption{Colour-colour diagram of all RXTE observations of 4U 1728--34. The gray crosses 
represent the data of the source from all available RXTE observations. Each point in this 
diagram corresponds to 256 s of data. We defined hard and soft colours as the 
$9.7-16.0/6.0-9.7$ keV and $3.5-6.0/2.0-3.5$ keV count rate ratios, respectively. 
The colours of 4U 1728--34 are normalised to the 
colours of Crab. The blue filled circles represent the colours of the persistent emission 
of the source at the onset of a PRE X-ray burst. The red star indicate the bursts with 
oscillations. PRE bursts with oscillations are therefore indicated with a red star inside a
blue circle. Black traigles represent bursts that show neither PRE nor oscillations.
The position of the source on the diagram is parametrized by the length 
of the black solid curve $S_{\rm a}$.}
    \label{fig ccd}
\end{figure*}

Coherent oscillations during bursts have been detected in many NS-LMXBs; these 
oscillations likely reflect the spin frequency of the NS \citep[e.g., ][]{Strohmayer97,Chakrabarty03}, 
but the mechanism of burst oscillations is still unclear \citep[e.g., ][]{Strohmayer96a, Muno04}. 
\cite{Strohmayer96b} suggested that burst oscillations are caused by asymmetries 
due to initially localized nuclear burning (the hot spot) that later spreads over the 
surface of the NS in the rise phase of the burst. This scenario has been supported
by \cite{Chakraborty14} using a large sample of bursts from 10 NS LMXB. 
This scenario, however, can not explain oscillations that persist for as long as 5$-$10 
s in the burst decay.

In some well-studied burst-rich accreting NS systems \citep[e.g. 4U 1636--53, 4U 1608--52, 
KS 1731--260 and Aquila X-1, ][]{Muno04, Galloway08a, Zhanggb11}, not all the bursts show 
coherent oscillations.  \cite{Muno04} showed that the appearance of burst oscillations 
is correlated with the accretion rate onto the NS. By analysing a large number 
of bursts in 4U 1636--53, \cite{zhanggb13} found that burst oscillations are always associated 
with an emitting area that remains constant for more than two seconds during the decaying phase 
of the burst. They suggested that tail oscillations could be due to the 
spread of a cooling wake,  which is formed by vortices during the cooling of the NS atmosphere
\citep{Spitkovsky02}. In this scenario, the speed of the cooling wake depends on the latitude 
at which the burst ignites, since the speed of the burning front near the equator is higher than that near 
the poles.  \cite{zhanggb13} suggested that in 4U 1636--53 the bursts with tail 
oscillations ignite at high latitude on the NS surface.
Rectntly \cite{Kajava14} found a correlation between the persistent spectral 
properties and the time evolution of the blackbody normalization during the burst 
decay. They suggested that this observed behaviour may be attributed to the accretion flow, 
which influences the spectrum of the cooling NS. It’s an interesting quesiton whether the link 
is between burst oscillations and time-resolved spectrum, or between accretion 
rate and burst oscillations and time-resolved spectrum. We investigate it here 
in another source $-$ 4U 1728--34.

\begin{figure*}
\renewcommand{\thefigure}{\arabic{figure}a}
    \centering
         \includegraphics[width=3.46in,height=2.5in,angle=0]{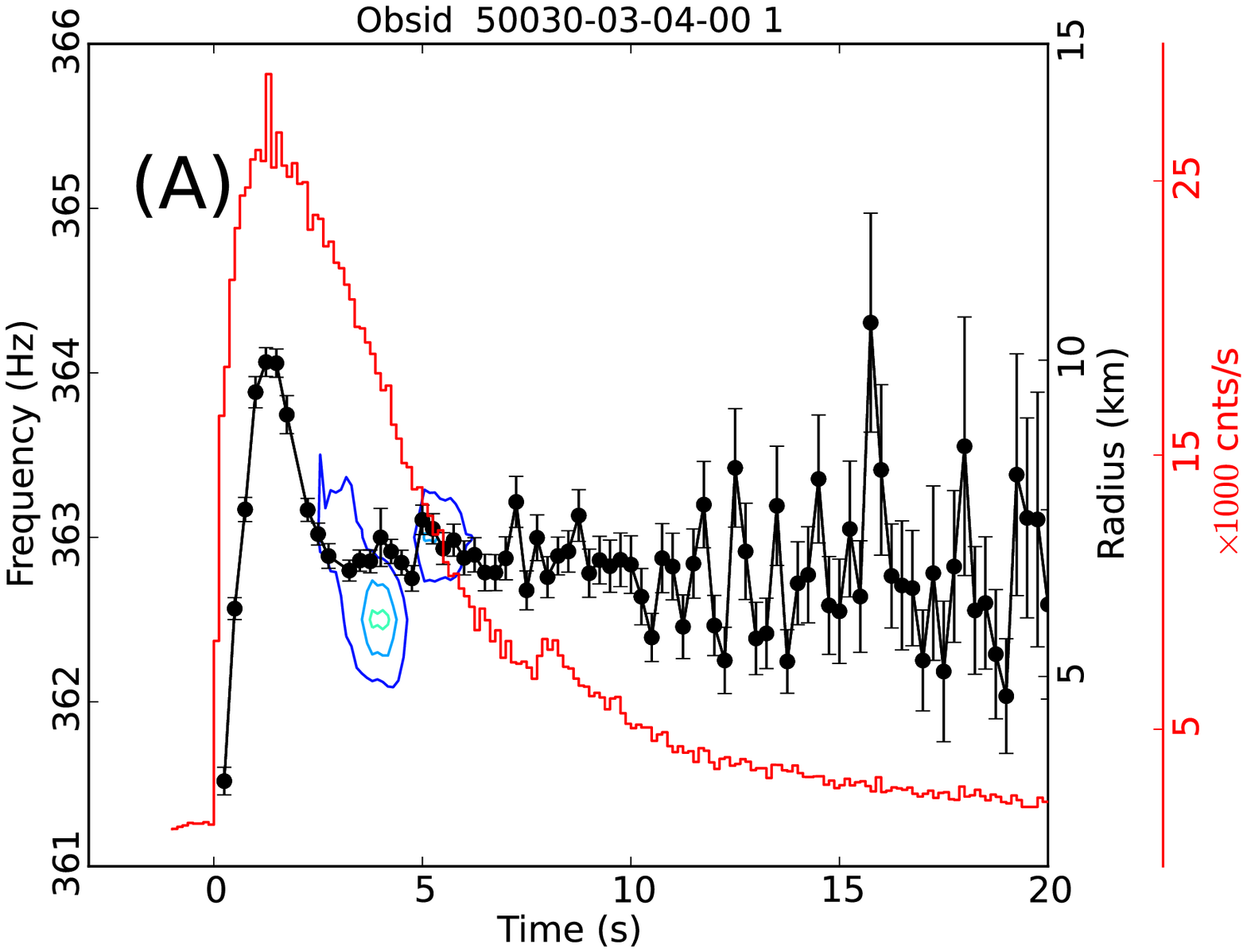}
         \includegraphics[width=3.46in,height=2.5in,angle=0]{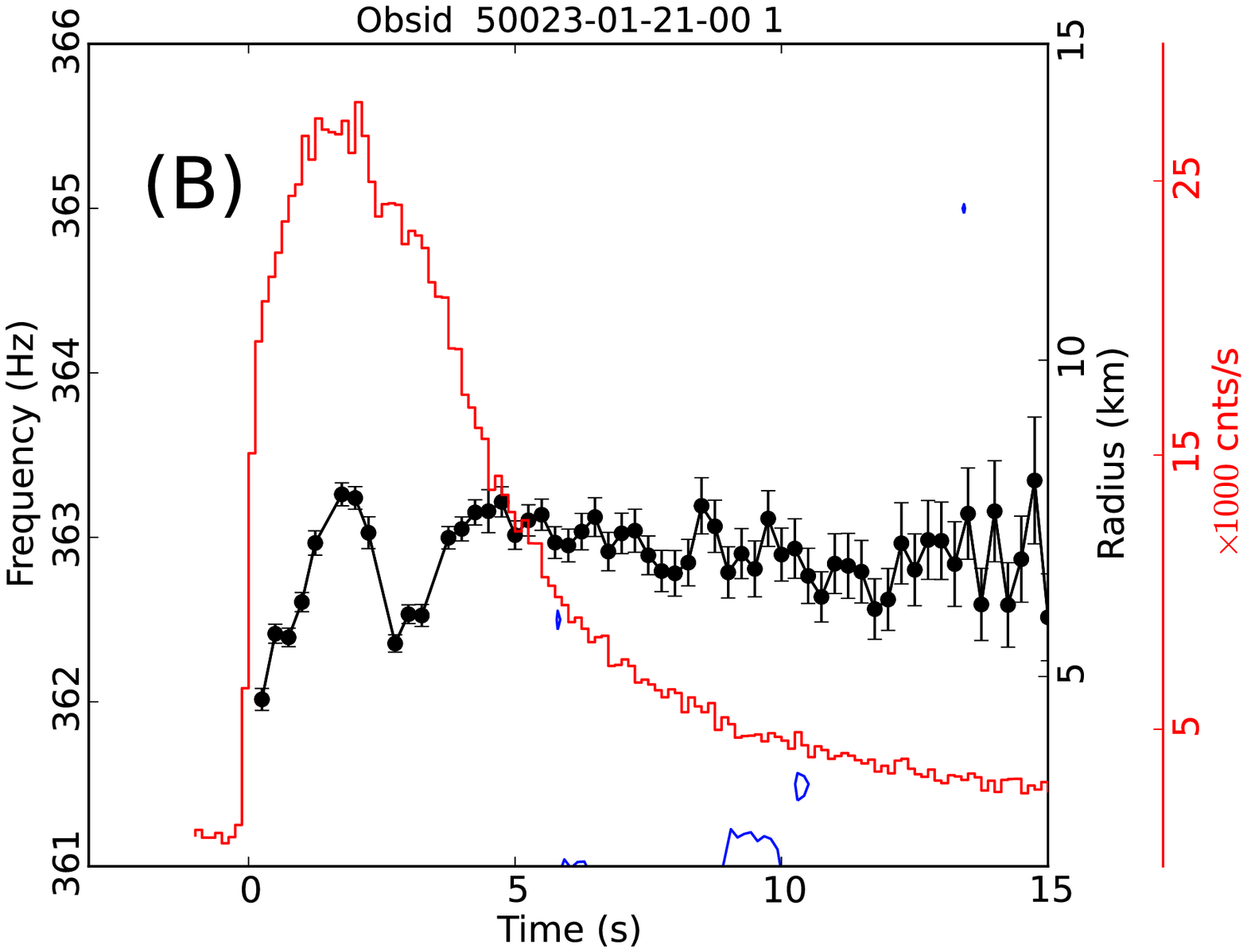}
         \includegraphics[width=3.46in,height=2.5in,angle=0]{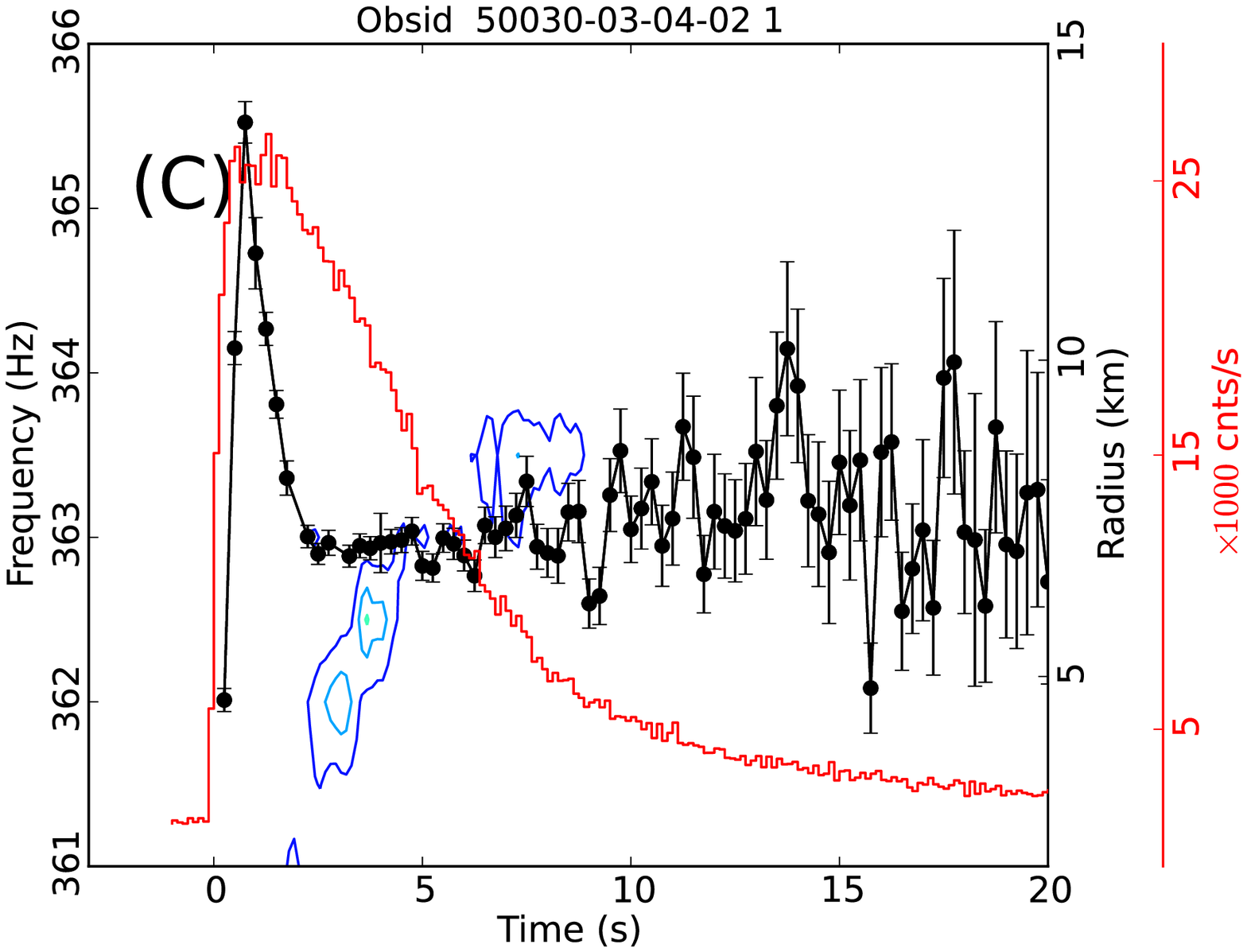}
         \includegraphics[width=3.46in,height=2.5in,angle=0]{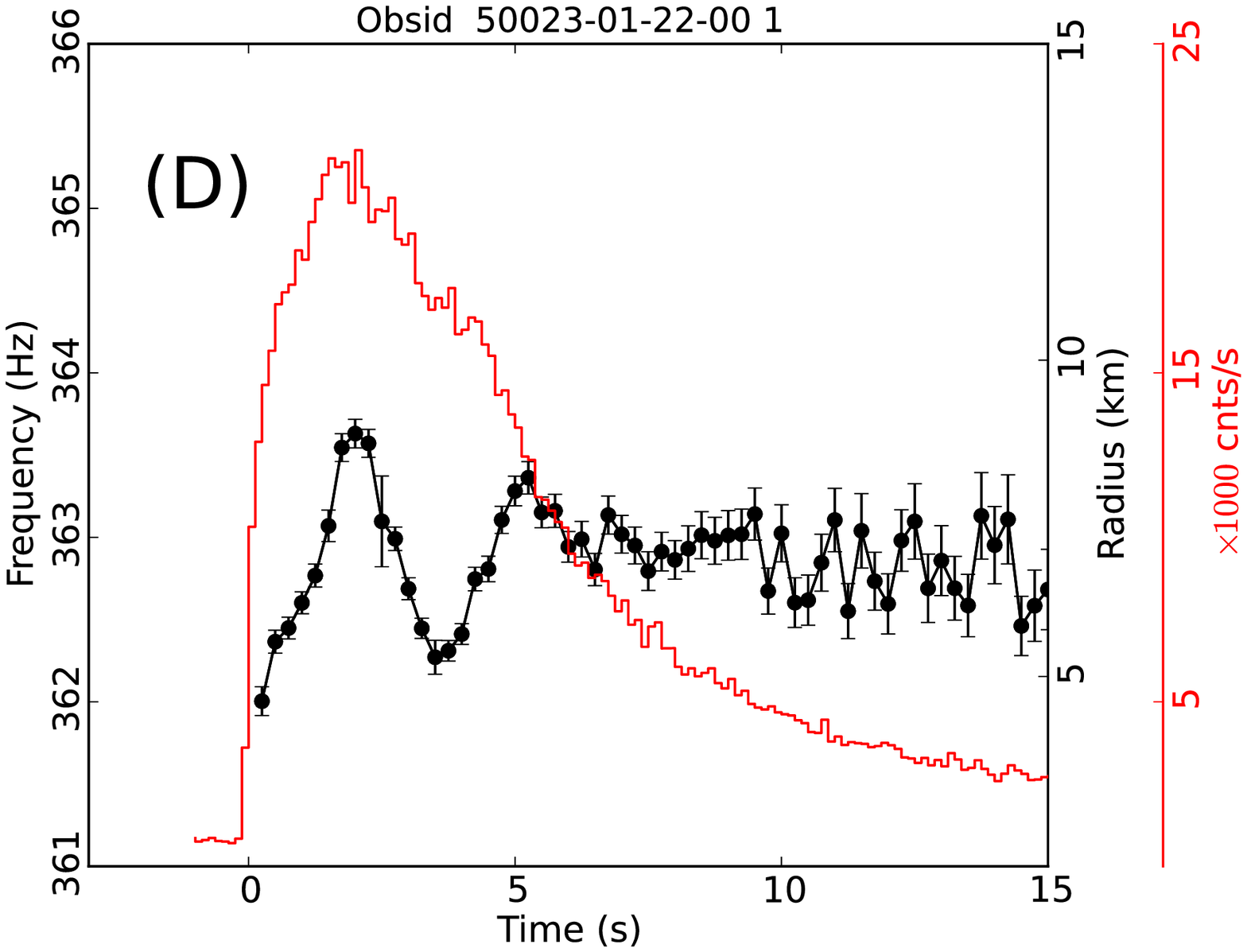}
         \includegraphics[width=3.46in,height=2.5in,angle=0]{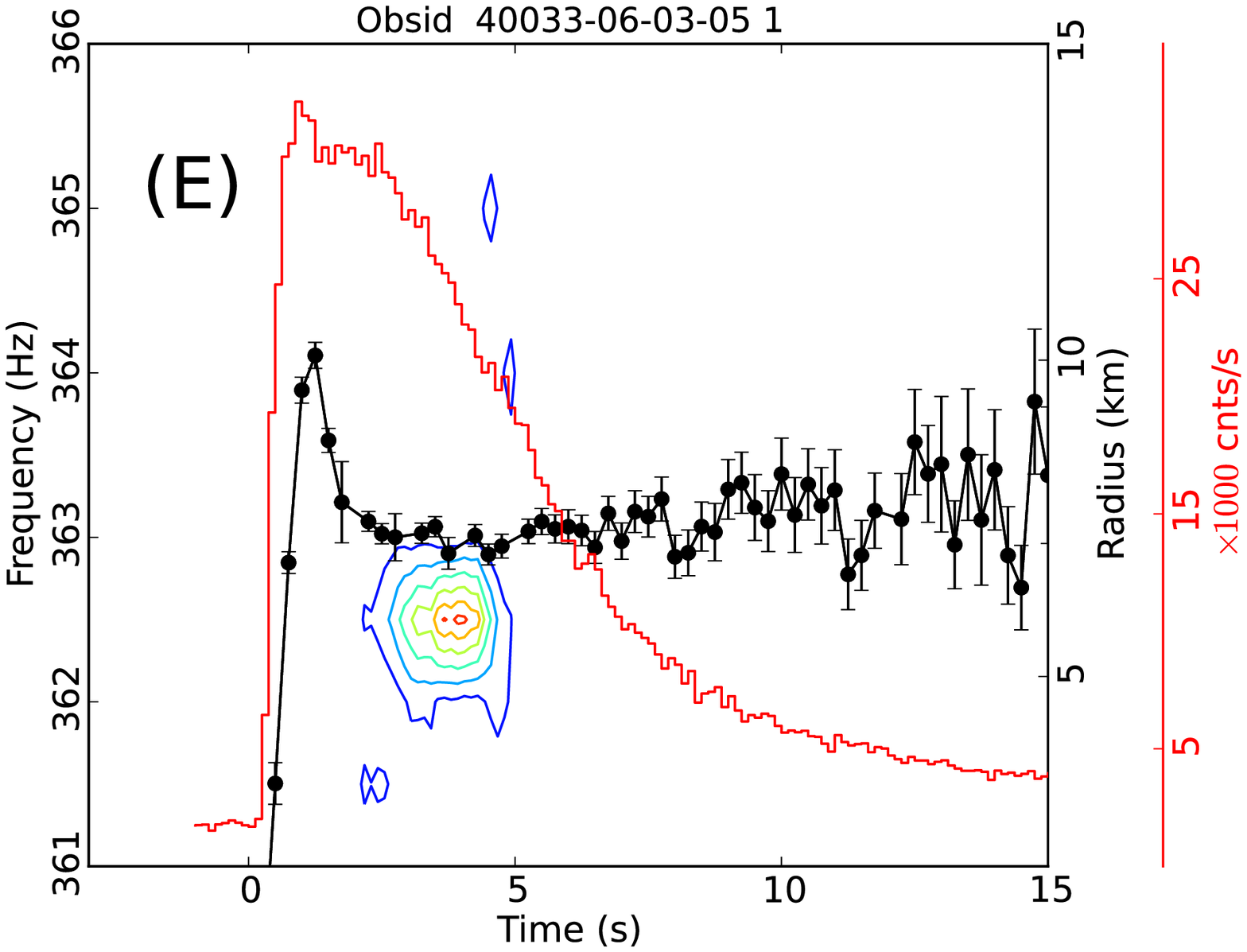}
         \includegraphics[width=3.46in,height=2.5in,angle=0]{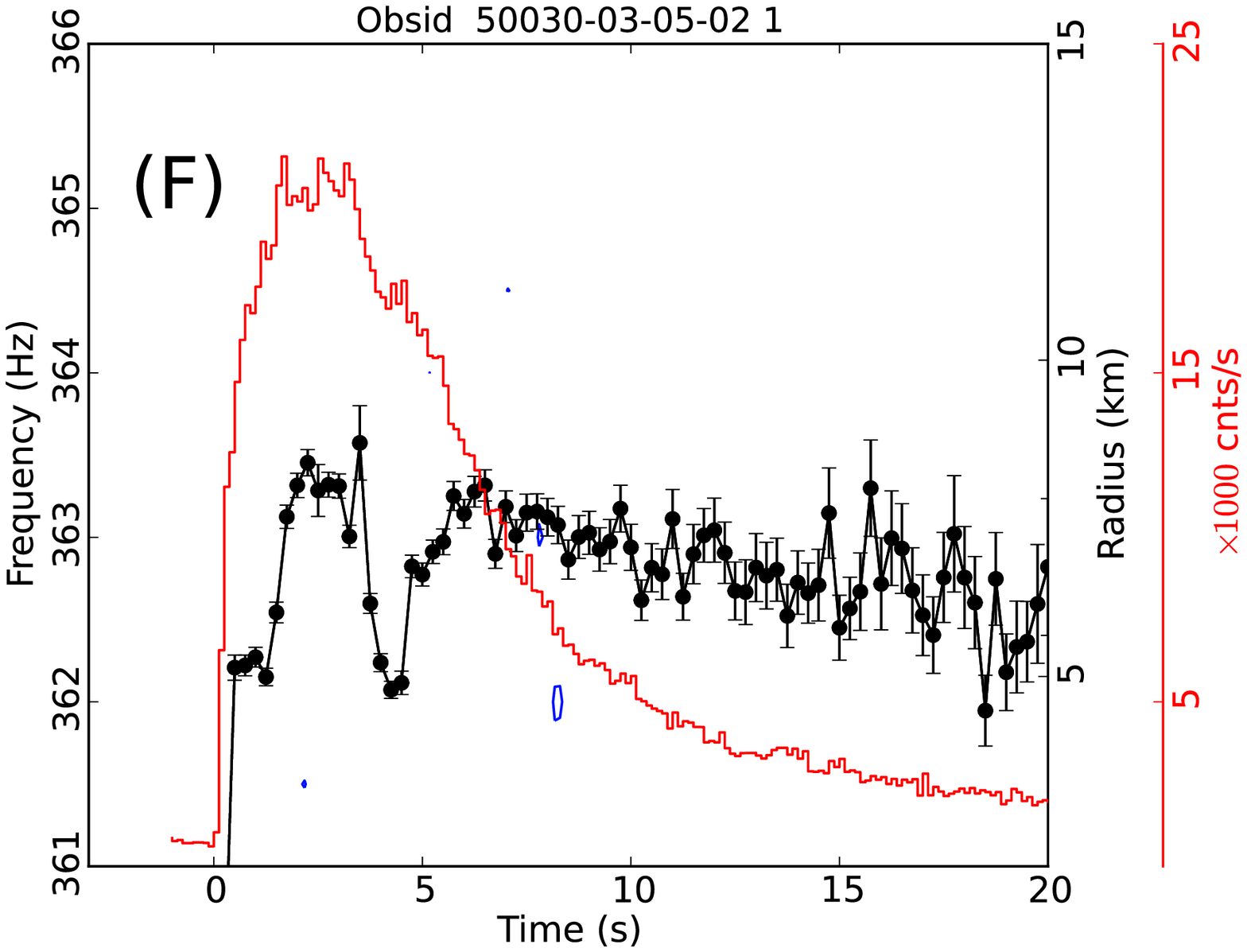}
\caption{ Left panels: burst with tail oscillations. Right panels: 
burst without tail oscillations. In each panel the red histogram shows the light curve of 
the burst at a resolution of 0.125 s. The intensity, in units of 1000 counts s$^{-1}$, 
is shown by the scale plotted to the right, outside of each panel. The contour lines show 
constant power values, increasing from 10 to 80 in steps of 10 (values are in Leahy units), 
as a function of time ($x$ axis) and frequency (left $y$ axis). The power spectra were 
calculated from 2 s intervals, with the start time of each successive interval shifted by 
0.125 s with respect to the start time of the previous interval. Black filled circles 
connected by a line show the best-fitting blackbody radius as a function of time at a 
resolution of 0.25 s (see the right $y$ axis), with error bars at the 90\% confidence level.
The burst light curve profile is aligned to the centre of each data interval used to 
calculate the power spectra and energy spectra. Note also the power contours at $\sim 363$ 
Hz during the bursts, which are due to oscillations in the burst. The label on top of 
each panel indicates the Obsid in which the burst was detected and the burst number in 
that observation.  All bursts in these
panels are PRE bursts.}
\label{fig_spectrum_1}
\end{figure*}

\begin{figure*}
\addtocounter{figure}{-1}   
\renewcommand{\thefigure}{\arabic{figure}b}
    \centering
         \includegraphics[width=3.46in,height=2.5in,angle=0]{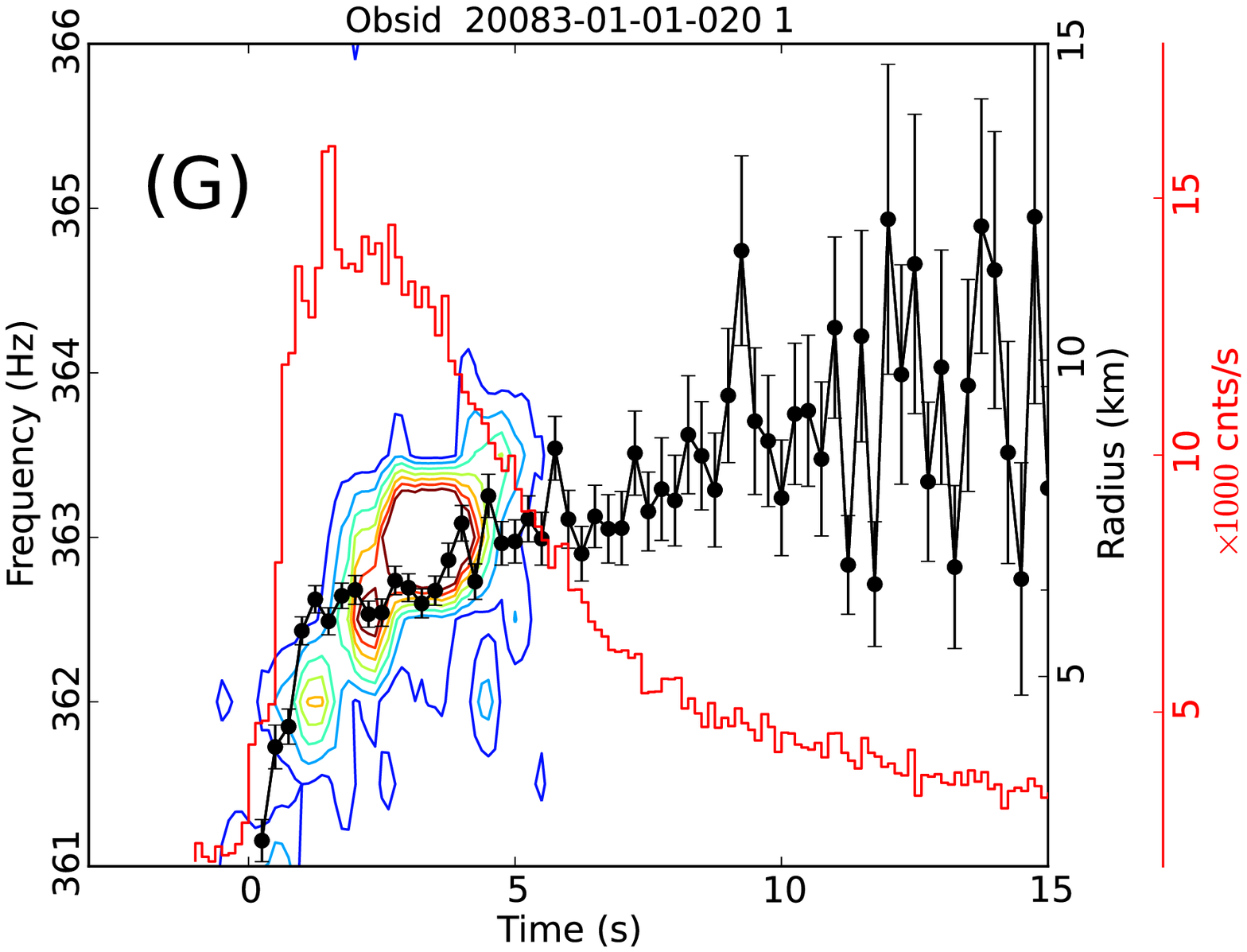}
         \includegraphics[width=3.46in,height=2.5in,angle=0]{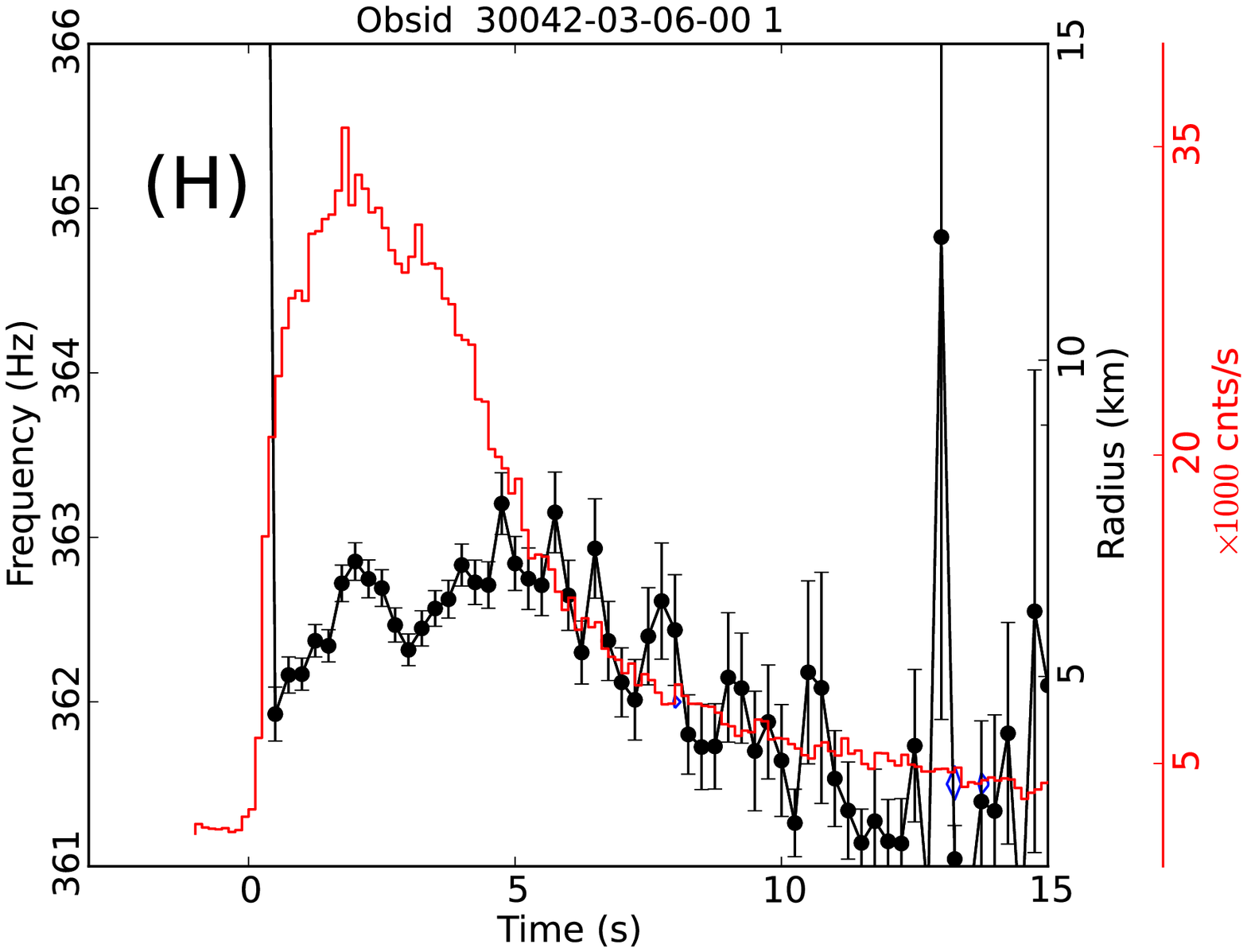}
         \includegraphics[width=3.46in,height=2.5in,angle=0]{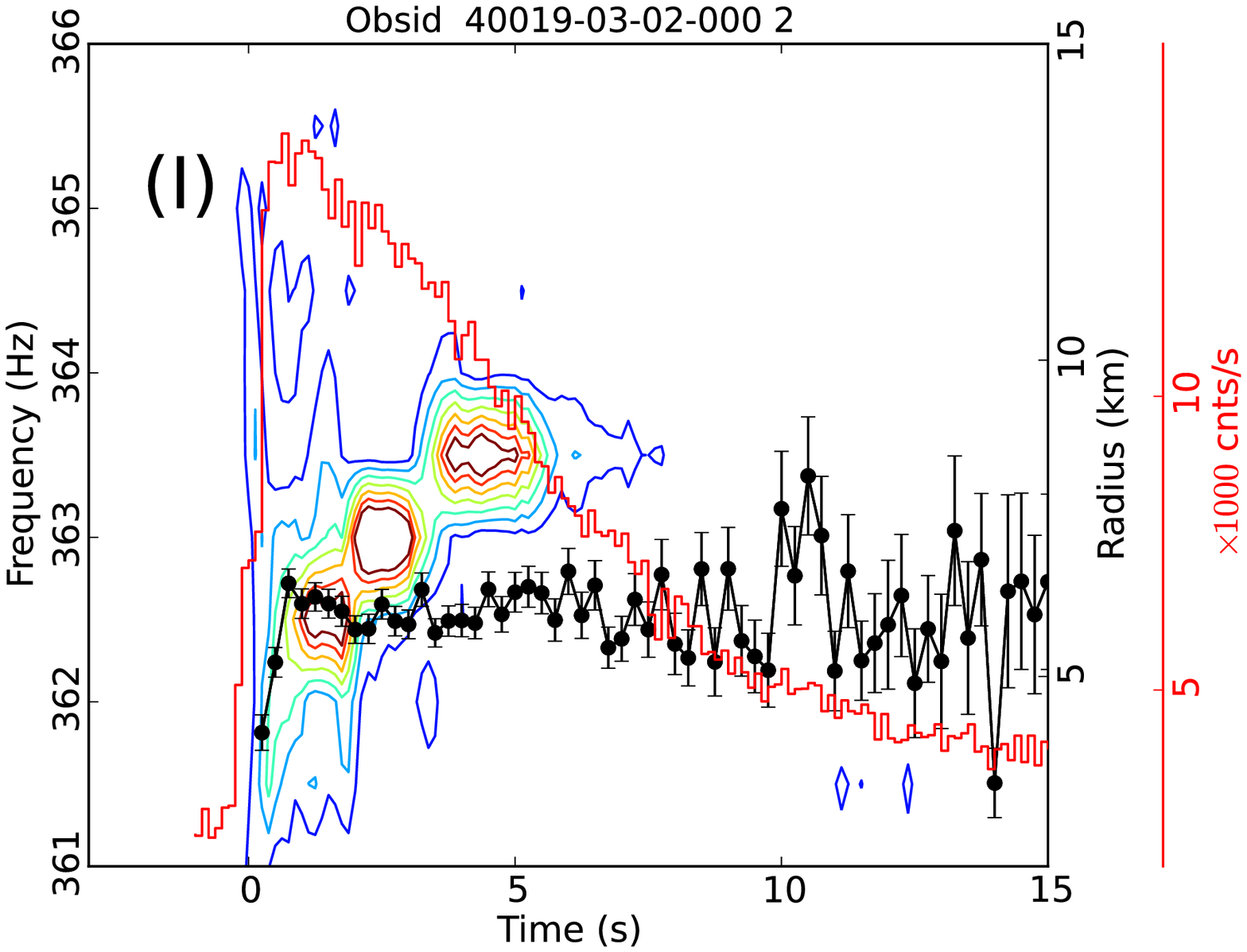}
         \includegraphics[width=3.46in,height=2.5in,angle=0]{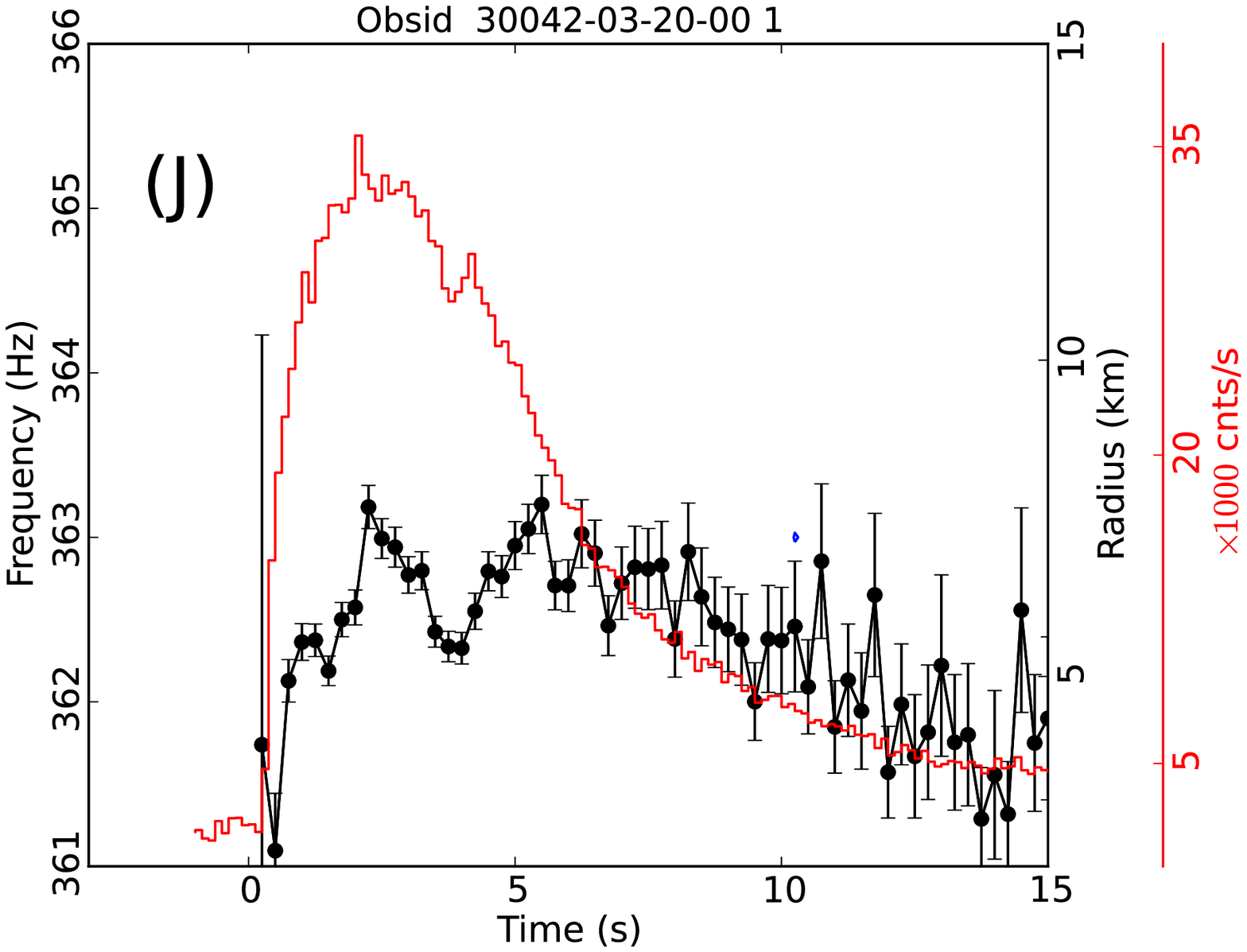}
         \includegraphics[width=3.46in,height=2.5in,angle=0]{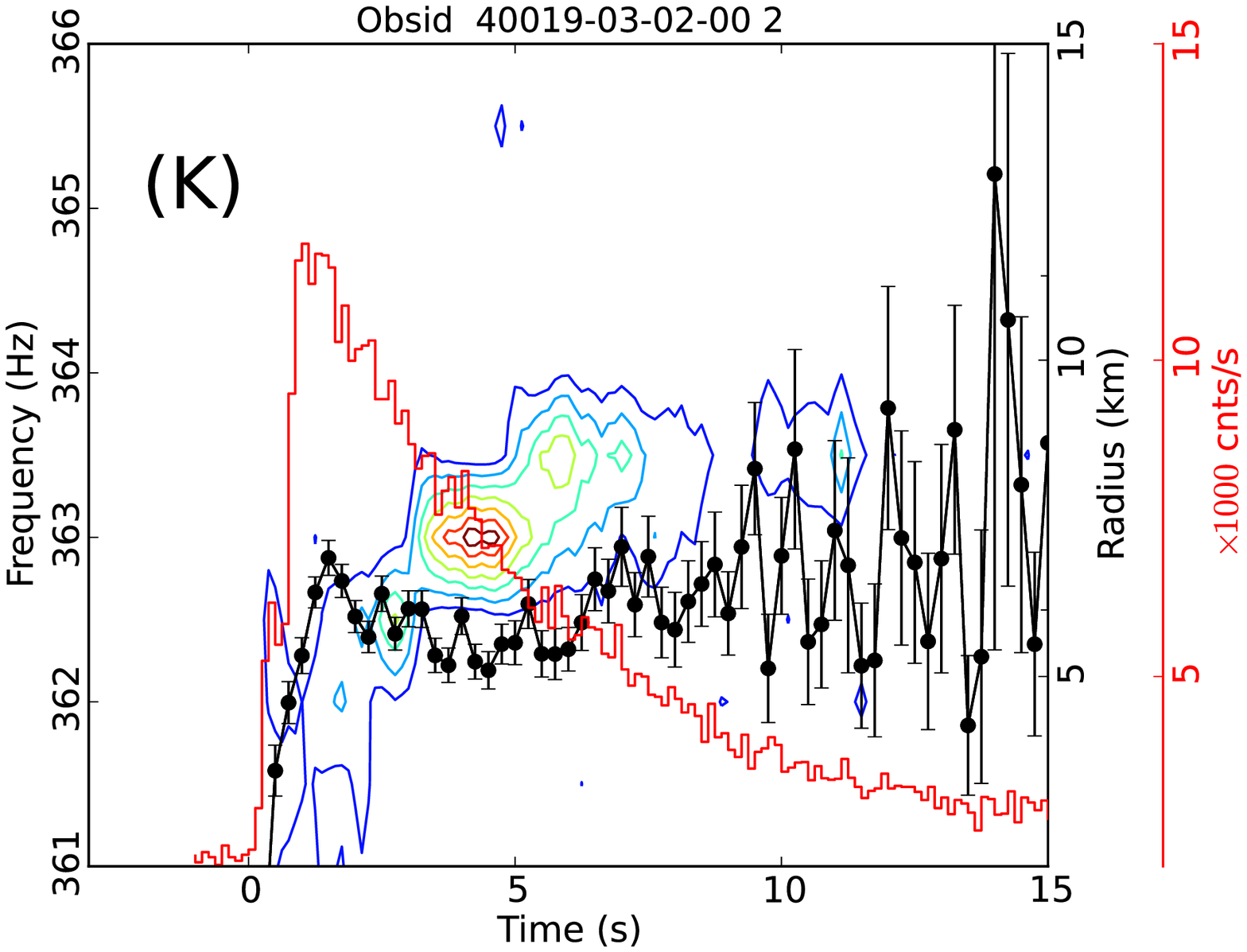}
         \includegraphics[width=3.46in,height=2.5in,angle=0]{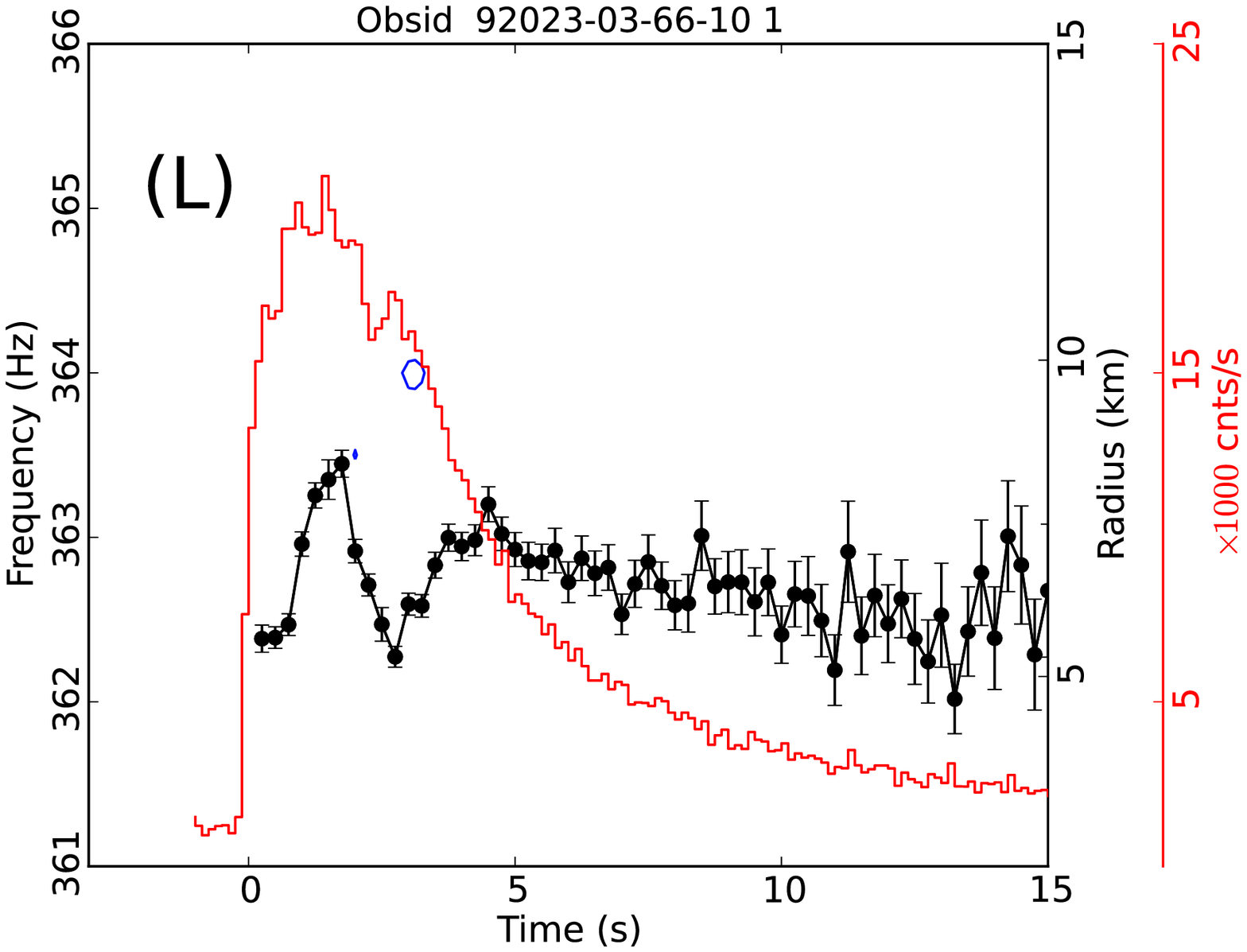}
    \caption{ Continued. Some of the bursts in these panels are non-PRE bursts.}
    \label{fig_spectrum_2}
\end{figure*}

Studying the burst morphology$-$rise shape$-$can also help us to understand 
the ignition latitude and other various processes operating during the burst rise.
\cite{Maurer08} developed a simple phenomenological model and they simulated burst 
light curves with ignition at different latitudes on the NS surface. 
They found that bursts that ignite near the equator 
always show convex rise shape, whereas bursts that ignite near the pole show 
either convex or concave rise shape.

The NS LMXB 4U 1728--34 is one of the best studied sources of type-I X-ray bursts 
\citep[e.g., ][]{Forman76, Hasinger89, Straaten01, Straaten02}. The distance to 
the source is between 4 and 6 kpc, based on the Eddington flux of the PRE bursts 
\citep[e.g., ][]{Disalvo00, Straaten01, Galloway08a}. The bursts in 4U 1728--34 have all 
short rise times and similar duration \citep{Straaten01, Galloway08a}, that is 
the characteristic behaviour of helium-rich bursts.  Using an analytical model 
of X-ray  burst spectral formation, \cite{Shaposhnikov03} claimed that 4U 1728--34 is 
an ultra-compact X-ray binary (UCXB) with an evolved, H-poor, donor.  Nearly coherent 
oscillations during X-ray bursts at 363 Hz from this source were first reported by 
\cite{Strohmayer96a}.  \cite{Straaten01}, \cite{Muno01} and \cite{Muno04} analysed 
observations of 4U 1728--34 with Rossi X-ray Timing Explorer (RXTE) and found a 
relationship between the appearance of burst oscillations  and the mass accretion rate.

In this paper we present the X-ray spectral and burst oscillation properties of 
all the type-I bursts in 4U 1728--34 observed with RXTE. In particular, we investigate 
the difference between the bursts with and  without oscillations. We find strong 
evidence for a link between coherent oscillations and the evolution of the apparent 
emission area during the bursts. We also find that the rise time and the shape of 
the light curve at the beginning of the bursts are different for bursts with and without 
oscillations in 4U 1728--34.  We describe the observations and data analysis 
in Section \ref{data}, and we present our results in Section \ref{result}. Finally, 
in Section \ref{discussion} we discuss our findings and  conclude in Section \ref{conclusion}

\section{Observation and data reduction}
\label{data}

We analysed all available data (411 observations for a total of  $\sim 1750$ ks) of 4U 1728--34 
from the Proportional Counter Array (PCA) on board RXTE \citep{Jahoda96, Jahoda06}. 
Due to the affect by the presence of the nearby transient 4U 1730--335 (the Rapid Burster),
12 observations were not used in this work.
The PCA consists of an array of five collimated proportional counter units (PCUs) operating 
in the 2$-$60 keV range. We produced 0.25-s light curves from the Standard-1 data (0.125-s 
time-resolution with no energy resolution) and searched for X-ray bursts in these light curves following 
the procedure described in \cite{Zhanggb11}.  We detected a total of 121 type-I X-ray 
bursts in these data.

We used the Standard-2 data (16-s time-resolution and 129 channels covering the full
$2-60$ keV PCA band) to calculate X-ray colours of the source. We defined
hard and soft colours as the $9.7-16.0/6.0-9.7$ keV and $3.5-6.0/2.0-3.5$ 
keV count rate ratios, respectively \citep[see][for details]{Zhanggb11}. We show the 
Colour-Colour Diagram (CD) of all 
observations of 4U 1728--34 in Figure \ref{fig ccd}. We parametrized the position of 
the source on the diagram by the length of the solid curve $S_{\rm a}$ 
\citep[see, e.g. ][]{Mendez99, Zhanggb11}, fixing the values of $S_{\rm a} = 1$ and 
$S_{\rm a} = 2$ at the top-right and the bottom-left vertex of the CD, respectively.

We used the high-time resolution modes that were available for each observation 
to produce  a spectrum every 0.25 s during  the whole duration of each burst. We generated the 
instrument response matrix for each spectrum with the standard {\sc ftools} routine 
{\sc pcarsp}, and we corrected each spectrum for dead time using the 
methods supplied by the RXTE team.  Due to the short exposure time used to 
generate each spectrum, statistical errors were dominant, and we did not include 
any additional systematic error to the spectra. For each burst we extracted the spectrum of the 
persistent emission just before or after the burst to use as background in our fits.

We fitted the spectra using {\sc xspec} version 12.8.0 \citep{Arnaud96}, restricting the 
spectral fits to the energy range $3.0 -20.0$ keV. We fitted the time-resolved net 
burst spectra with a single-temperature blackbody model ({\sc bbodyrad} in {\sc xspec}), 
as generally burst spectra are well fitted by a blackbody \citep[e.g., ][]{Straaten01, Kuulkers02, 
Galloway08a}. We also included the effect of interstellar absorption along the line of sight using the 
{\sc xspec} model $wabs$. During the fitting we kept the hydrogen column density, $N_{\rm H}$, 
fixed at  $2.3\times10^{22} $cm$^{-2}$ \citep{DA06}, and to calculate the radius of the 
emitting blackbody area in km, $R_{\rm bb}$, we assumed a distance of 5.2 kpc 
\citep[e.g., ][]{Disalvo00, Straaten01, Galloway08a}.

For each burst we computed Fourier power density spectra (PDS) from 2-s data segments 
for the duration of the burst using the 125 $\mu$s binned data over the full PCA band pass, 
setting the start time of each segment to 0.125 s after the start time of the previous segment. 
We used these PDS to produce time-frequency plots \citep[also known 
as dynamic power spectra; see][]{Berger96} for each burst.  For coherent oscillations, 
we only searched the frequency range $360 - 365$ Hz with a resolution of 0.5 Hz. 
We considered that a signal was significant if it had a probability of
$< 10^{-4}$ that it was produced by noise accounting for the number of possible 
independent trials, and if the signal appeared in at least two PDS within the 
tail of a single burst. \citep[See discussion in][for details about the detection 
and measurement of burst oscillations.]{zhanggb13}

\begin{figure*}
\renewcommand{\thefigure}{\arabic{figure}a}
    \centering
         \includegraphics[width=3.45in,angle=0]{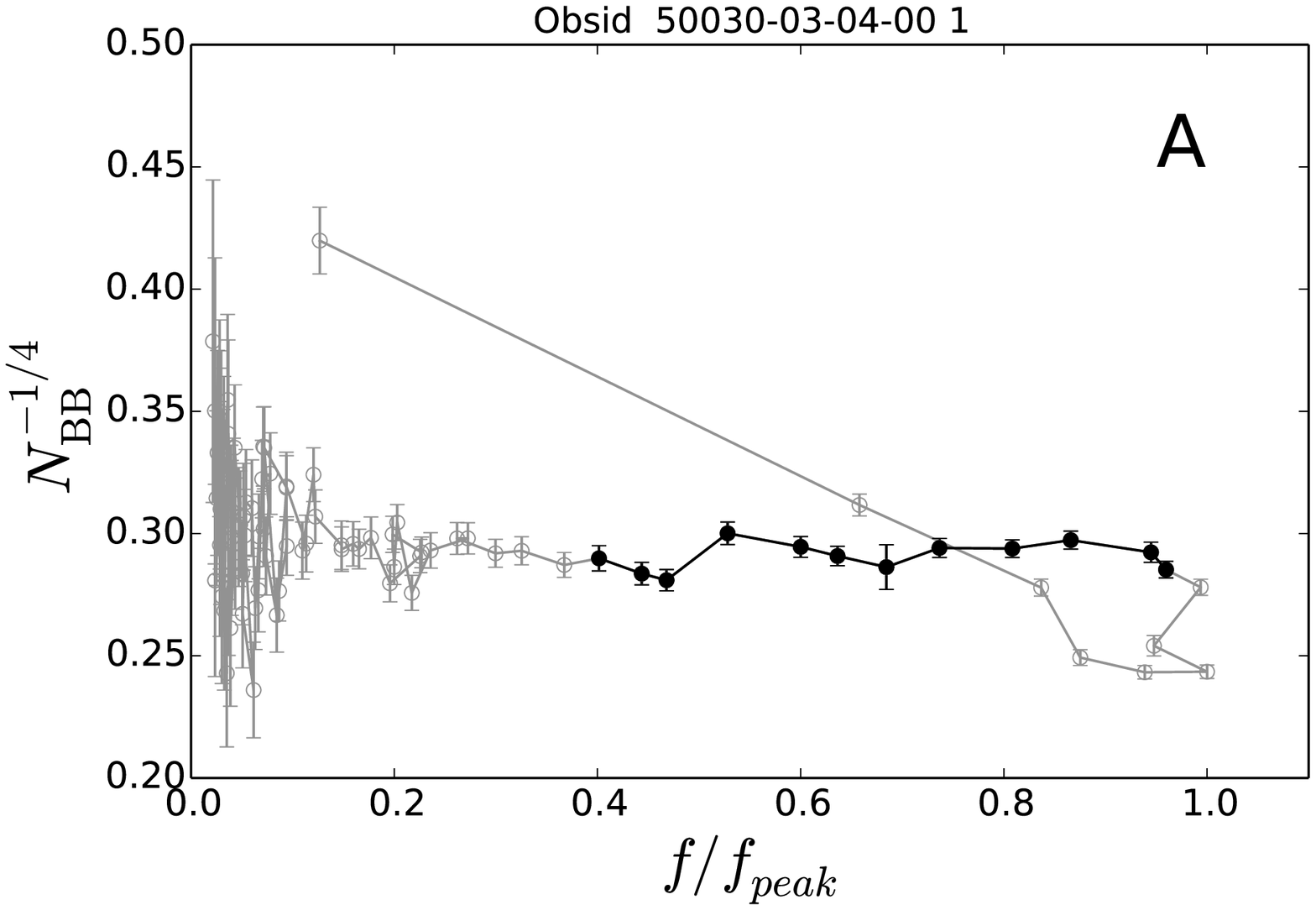}
         \includegraphics[width=3.45in,angle=0]{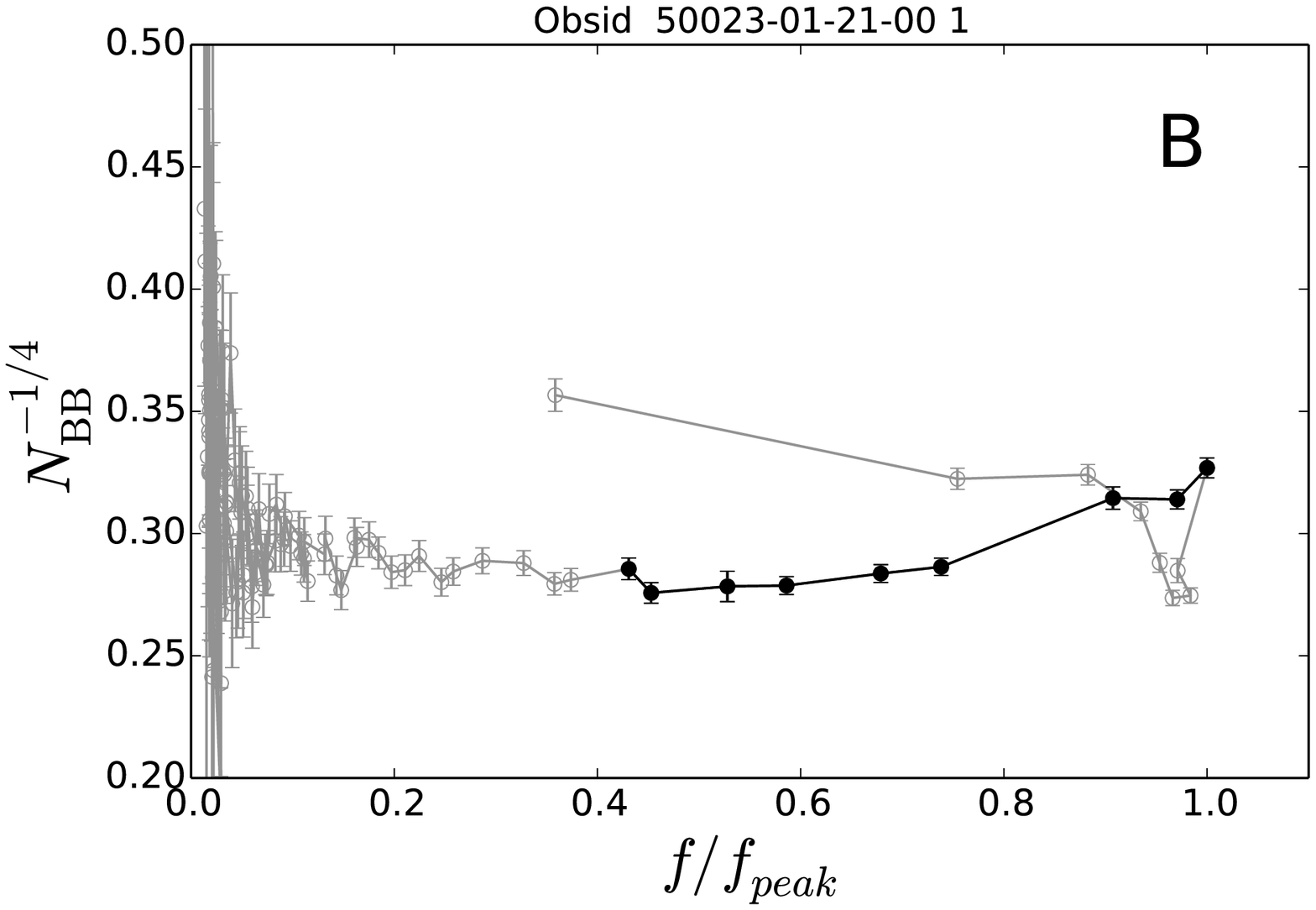}
         \includegraphics[width=3.45in,angle=0]{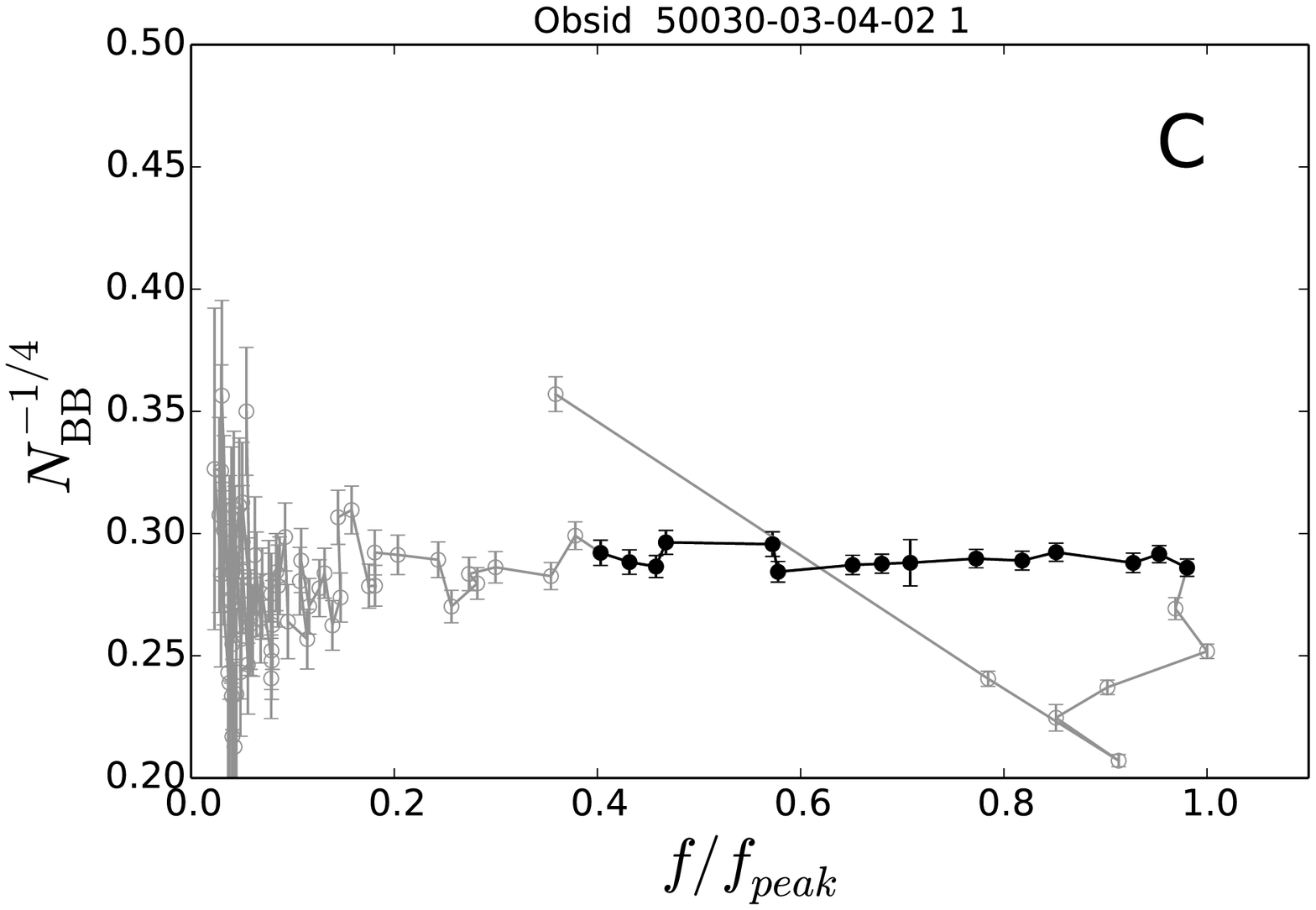}
         \includegraphics[width=3.45in,angle=0]{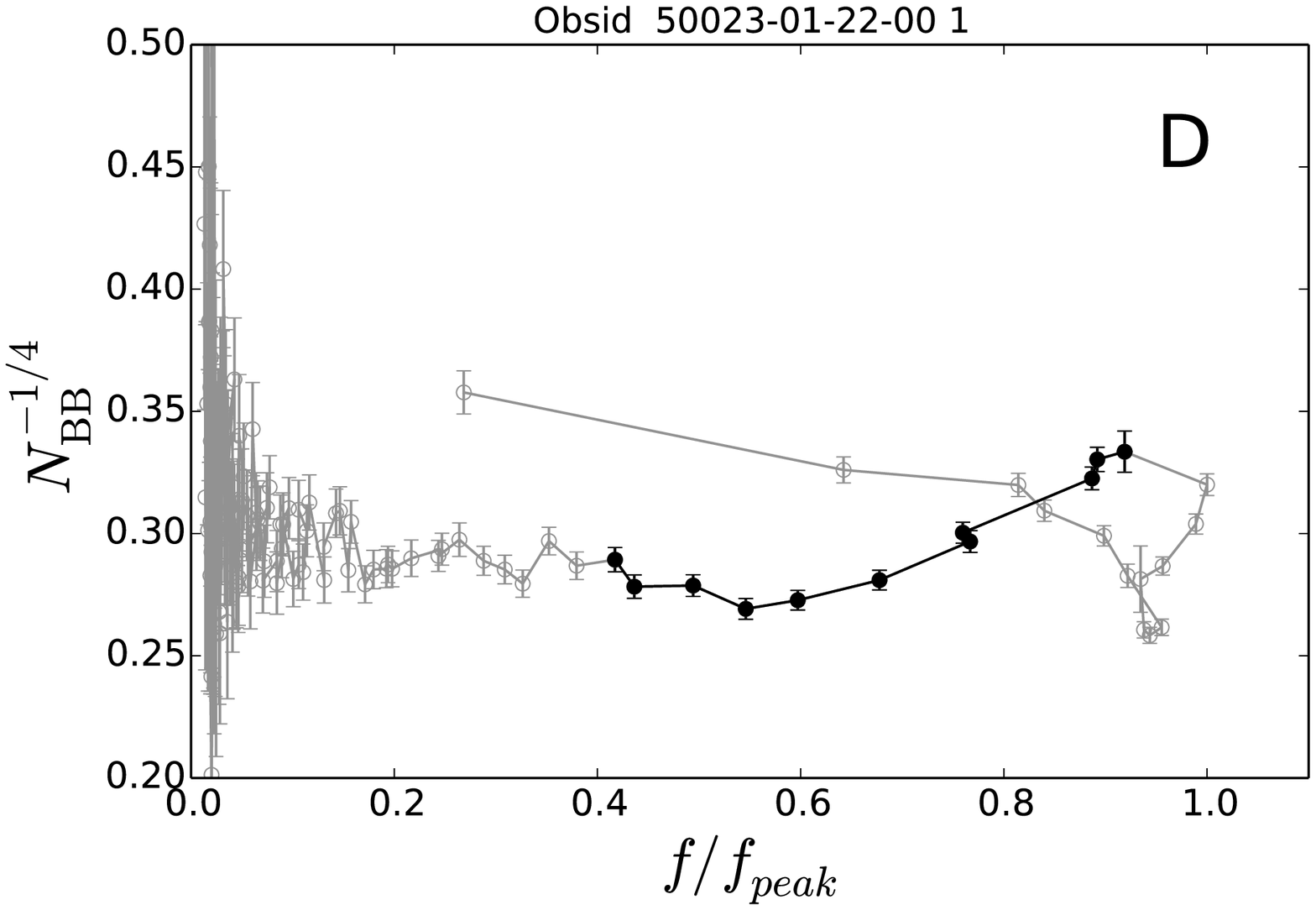}
         \includegraphics[width=3.45in,angle=0]{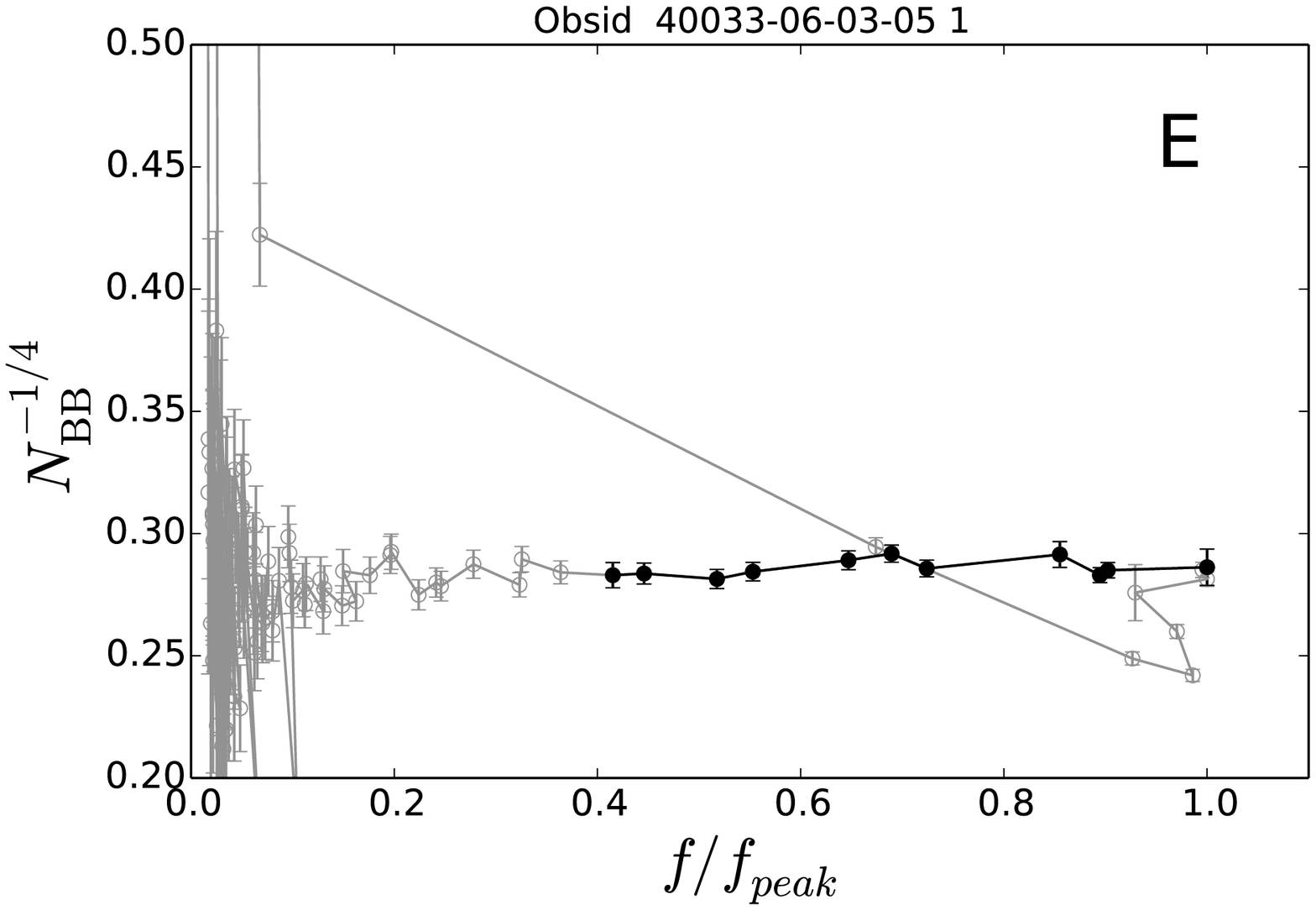}
         \includegraphics[width=3.45in,angle=0]{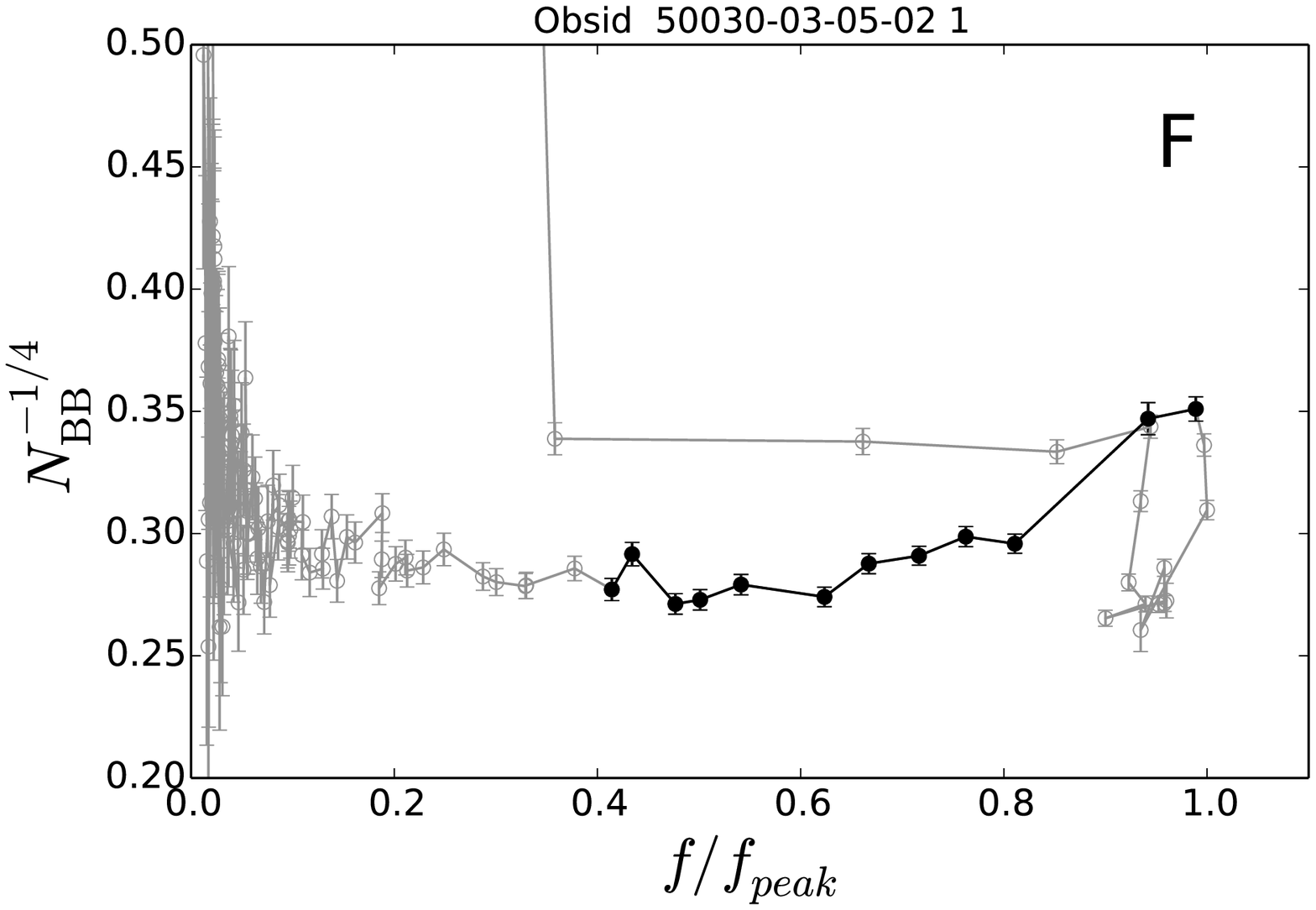}
    \caption{The $N_{BB}^{–1/4}$ as a function of flux during the PRE type-I X-ray burst 
from 4U 1728--34 shown in Figure \ref{fig_spectrum_1}. The black filled circles correspond to the 
time intervals in the decaying phase of the bursts in which the flux of the burst is 
between 40\% and 100\% of, respectively, the touch down flux for PRE bursts or the peak 
flux for non-PRE bursts. The open gray circles correspond to the data outside those intervals.
The error-bars of $N_{BB}^{-1/4}$ represent the $1 \sigma$ confidence level. }
    \label{fig_flux_fc1}
\end{figure*}

\begin{figure*}
\addtocounter{figure}{-1}    
\renewcommand{\thefigure}{\arabic{figure}b}
    \centering
         \includegraphics[width=3.45in,angle=0]{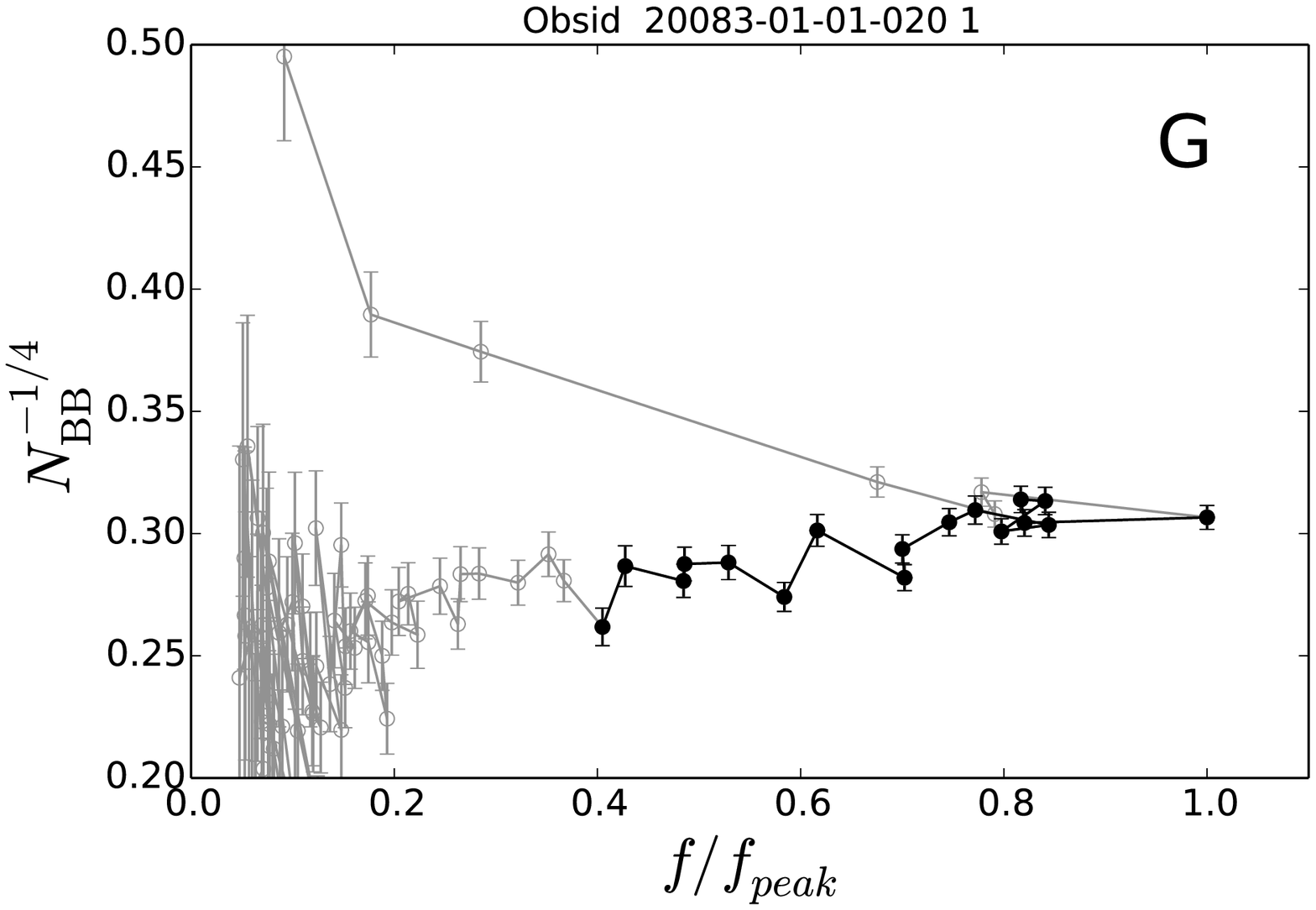}
         \includegraphics[width=3.45in,angle=0]{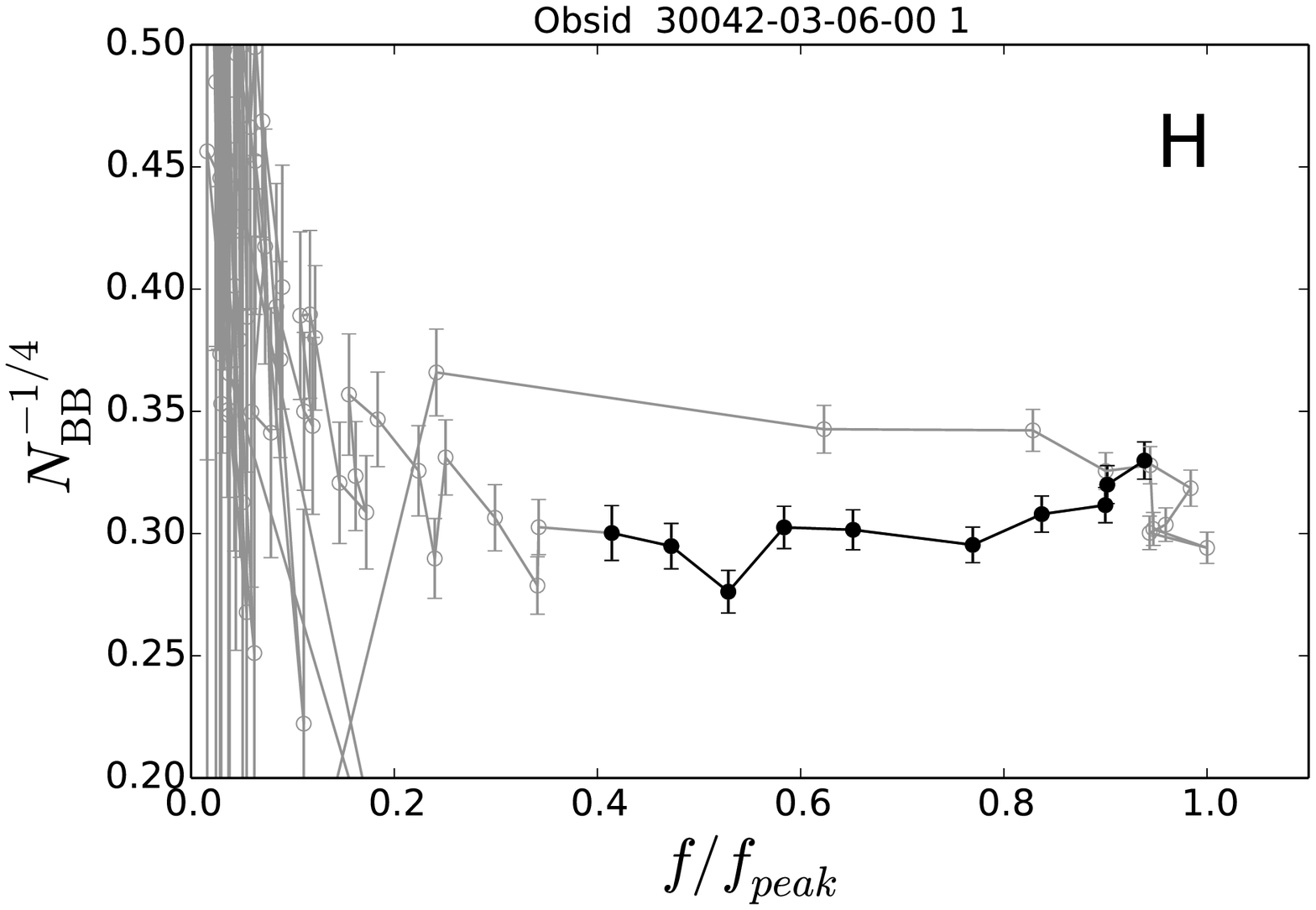}
         \includegraphics[width=3.45in,angle=0]{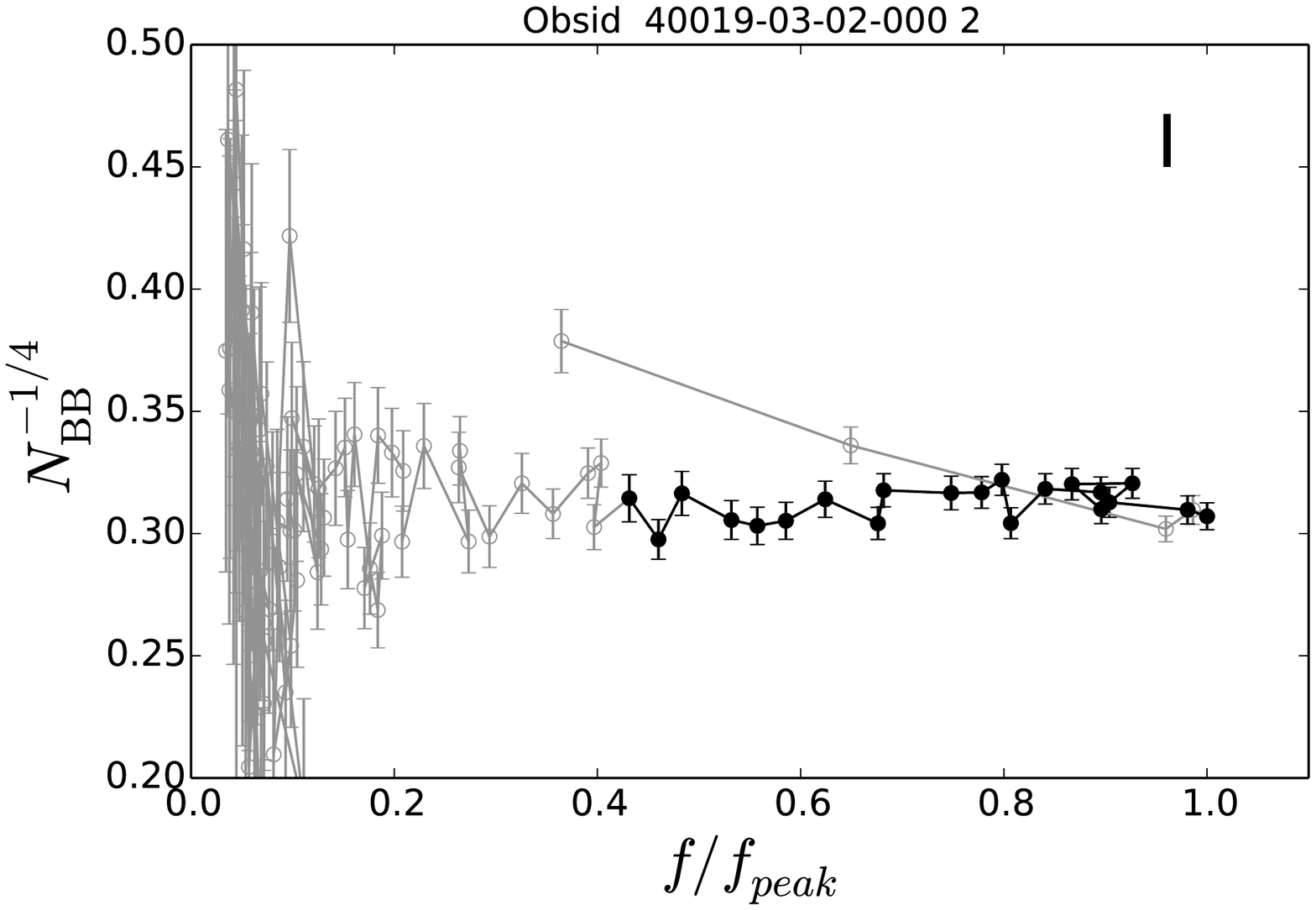}
         \includegraphics[width=3.45in,angle=0]{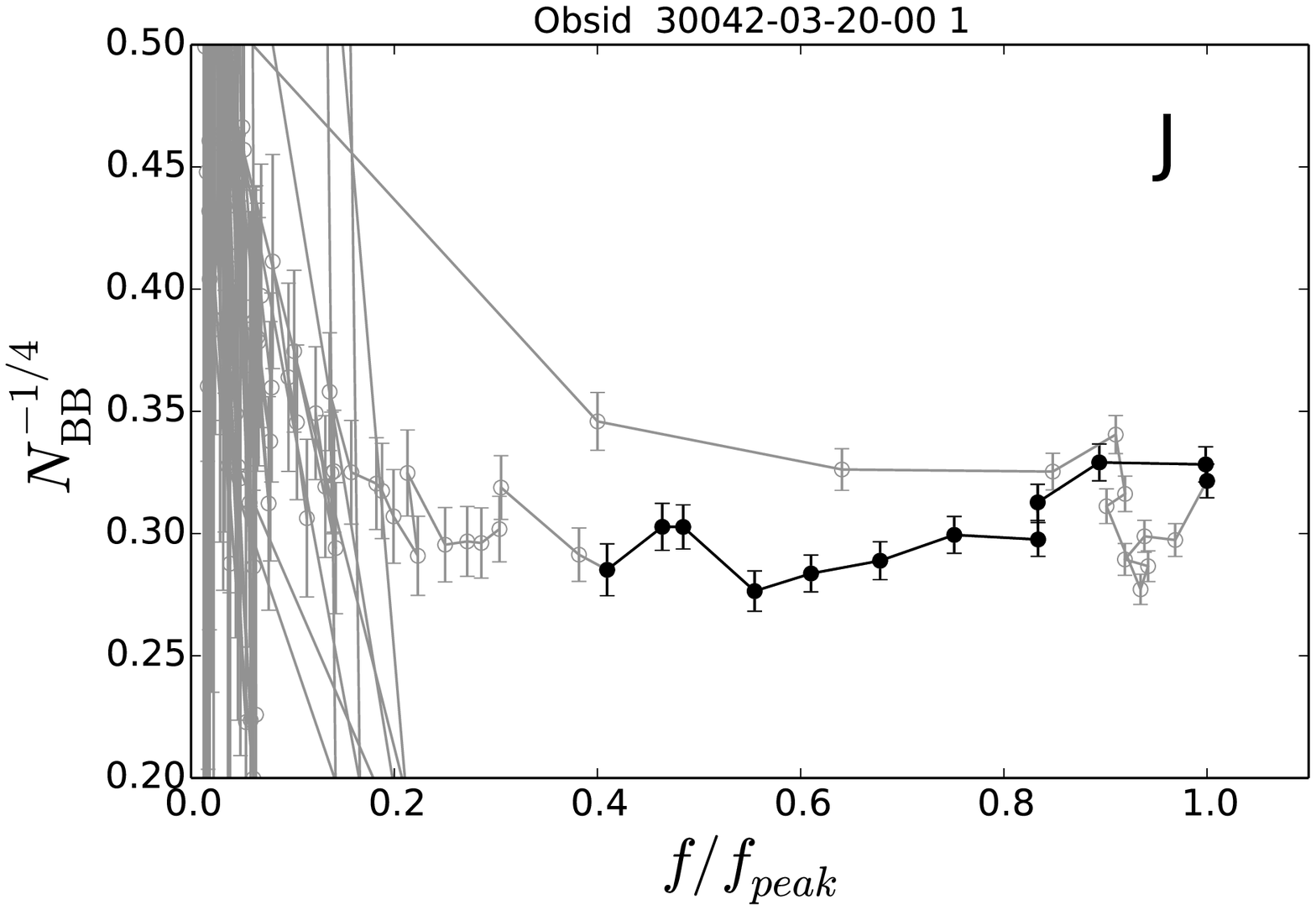}
         \includegraphics[width=3.45in,angle=0]{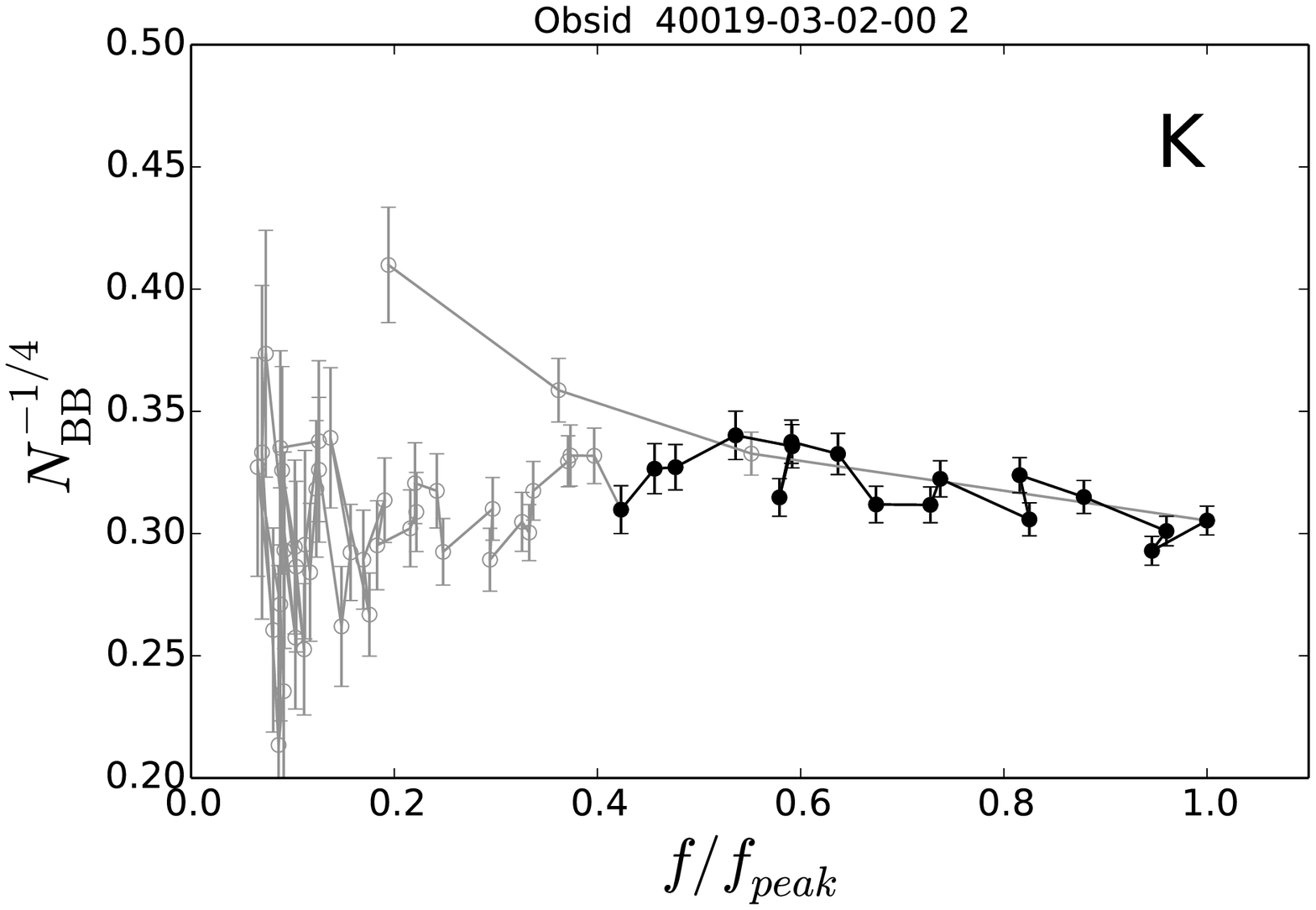}
         \includegraphics[width=3.45in,angle=0]{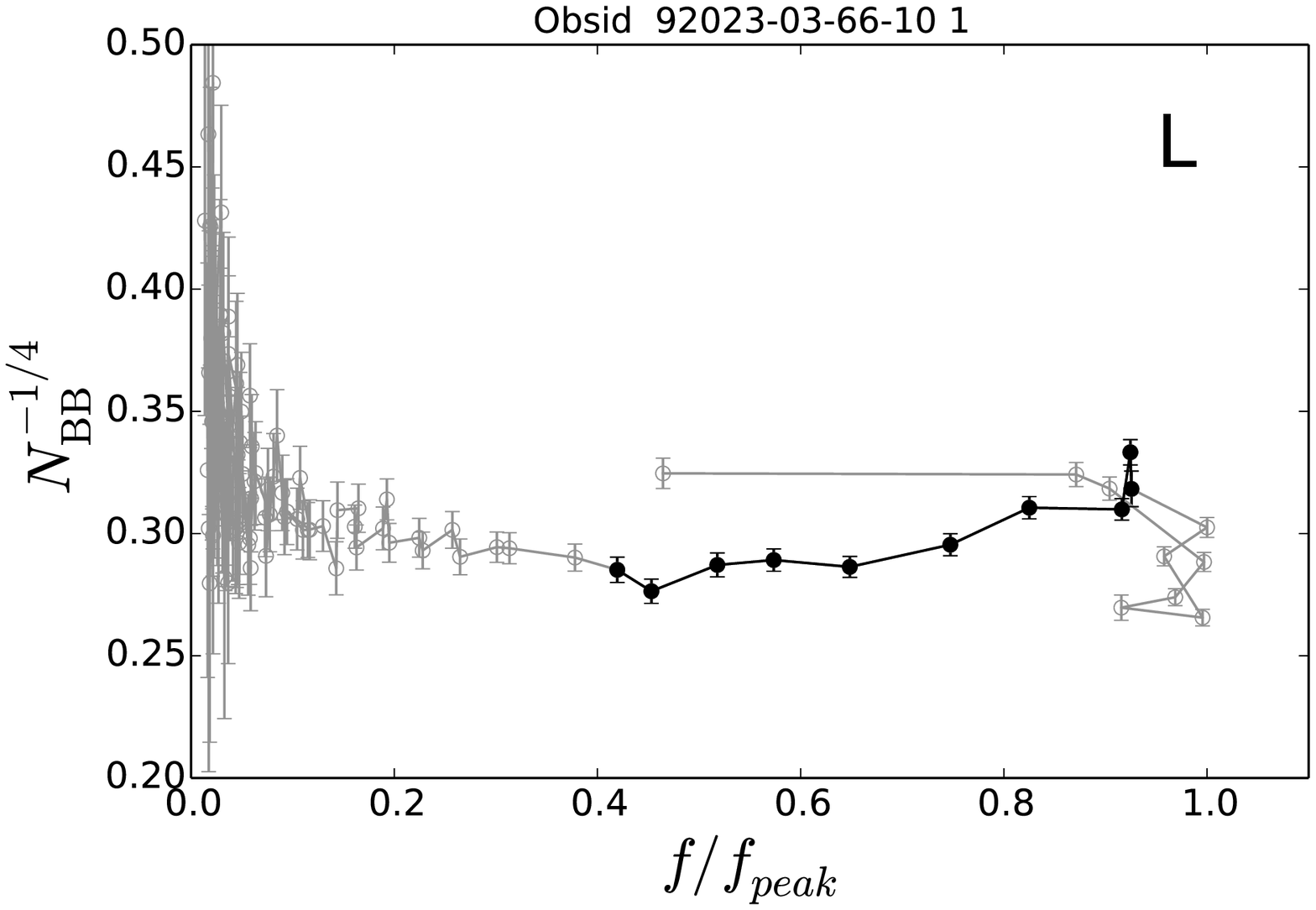}
    \caption{ Continued. Same as in Figure \ref{fig_flux_fc1} for the bursts in Figure 
    \ref{fig_spectrum_2}. Some of these are non-PRE bursts. }
    \label{fig_flux_fc2}
\end{figure*}

\section{Results}
\label{result}

\subsection{Time-resolved spectra and burst oscillations}
\label{time-resolved spectra}

In Figure \ref{fig_spectrum_1} and \ref{fig_spectrum_2} we show twelve examples of type-I 
X-ray bursts of 4U 1728--34. In the left panels of this Figure we show six cases of bursts 
with oscillations, while in the right panels we show six cases of bursts without oscillations. 
In each panel the red histogram shows the shape of the light curve of the bursts at a 
resolution of 0.125 s. The contour lines show constant Fourier power values, increasing from 
10 to 80 in steps of 10 (values are in Leahy units), as a function of time ($x$-axis) and 
frequency (left $y$-axis). Black filled circles connected by a line show the fitted blackbody 
radius as a function of time at 0.25-s time resolution (see the right $y$-axis). The burst 
light curve is aligned to the centre of each time interval used to calculate the power and energy spectra.

In the PRE bursts of 4U 1728--34, Figure \ref{fig_spectrum_1} shows that the behaviour of the 
blackbody radius during the photospheric radius expansion phase is quite different 
between bursts with and without oscillations. In the PRE bursts with oscillations
the blackbody radius increases very fast ($\sim 1$ s) to a maximum value and then 
decreases, whereas in the PRE bursts without oscillations the blackbody radius 
increases somewhat slower ($\sim 2$ s to reach the maximum value).

We also find that the behaviour of the blackbody radius after the touch down point is not the same for
all the PRE bursts in 4U 1728--34. In PRE bursts with oscillations, 
after the expansion phase the blackbody radius, $R_{\rm bb}$, first decreases rapidly, 
it then stays constant for a while, and finally it either increases slightly, or it 
remains more or less constant towards the tail of the burst (see left panels of Figure 
\ref{fig_spectrum_1}). In the PRE bursts without oscillations, after the expansion phase the 
blackbody radius first decreases rapidly to a  minimum, then it immediately increases again 
very quickly, and finally it either decreases slightly, or it remains more or less constant 
(see right panel of Figure \ref{fig_spectrum_1}). This behaviour is similar to the one found in 
the LMXB 4U 1636--53  by \cite{zhanggb13}.

We detected nine non-PRE bursts with tail oscillations in our observations of 4U 1728--34. 
Similar to the case of PRE bursts in this source, after the peak of the burst these 
non-PRE bursts show a period in which $R_{\rm bb}$ remains 
more or less constant during the time in which tail oscillations are present (see 
the left panels of Figure \ref{fig_spectrum_2}).

In \cite{zhanggb13} we found that, for PRE bursts, in 4U 1636--53 the presence of burst 
oscillations is associated with the duration of the post touchdown phase. This distinction,
however, only works for 
PRE bursts, since there is no touch down or post touch down phase in non-PRE bursts. 
In the following subsections we explore other properties of the bursts that, at least 
in the case of 4U 1728-34, appear to be associated with the presence of burst oscillations 
in the tail of the bursts.

\subsection{The blackbody normalization as a function of flux}
\label{colour factor}

From blackbody fits to the time-resolved spectra of the thermonuclear X-ray bursts we 
can obtain $R_{\rm bb}$ and $T_{\rm bb}$, the blackbody radius and colour temperature, 
respectively, from which we can calculate the bolometric flux of the neutron star. 
\cite{Zhanggb11} and  \cite{Garcia13} found that, in the cooling phase of bursts in 
4U 1636--53 and 4U 1820-30, the relation between the bolometric flux and the temperature 
is very different from the canonical $F_{\rm b} \propto T^4_{\rm bb}$ relation which 
is expected if the apparent emitting area on the surface of the NS remains constant 
and if the spectrum is blackbody. Departures from the $F_{\rm b} \propto T^4_{\rm bb}$ 
relation have also been reported from other NS-LMXB systems \citep{Suleimanov11, Kajava14}. 
This could be due to either changes in the emitting area of the neutron star during 
this phase, or to changes in the colour-correction factor,
$f_{\rm c} = T_{\rm bb}/T_{\rm eff} = \sqrt{R_{\infty}/R_{\rm bb}}$, 
where $R_{\infty}$ is the NS radius observed at infinity and $T_{\rm eff}$ is the 
effective surface temperature.  If the apparent emitting area 
on the surface of the NS remains constant as the flux decreases during the decay of the 
bursts, $f_{\rm c}$ can be expressed in terms of the  
blackbody normalization: $N_{BB}^{-1/4} = f_{\rm c} \times A$, where 
$\qquad A = (R_{\infty } [{\rm km}]/D_{10})^{-1/2}$, with $D_{10}$ the distance to the 
source in units of 10 kpc \citep{Suleimanov11}.

In Figures \ref{fig_flux_fc1} and \ref{fig_flux_fc2} we show  $N_{BB}^{-1/4}$  as a 
function of flux during the same 12 X-ray bursts shown in Figures \ref{fig_spectrum_1}
and \ref{fig_spectrum_2}. The flux in the $x-$axis is normalised by the peak flux of 
each burst individually. In the left panels of this Figures we show bursts with oscillations,  
while in the right panels we show bursts without oscillations. 
We found that in bursts with oscillations, especially in the case of PRE bursts 
(see Figures \ref{fig_flux_fc1}), at the beginning of the cooling phase and immediately after 
the touch down point, the value of  $N_{BB}^{-1/4}$ remains more or less constant 
as the flux decreases, whereas in bursts without oscillations the value of  $N_{BB}^{-1/4}$
decreases when the flux decreases in the decay phase of the bursts.

\begin{figure}
    \centering
        \includegraphics[width=3.25in,angle=0]{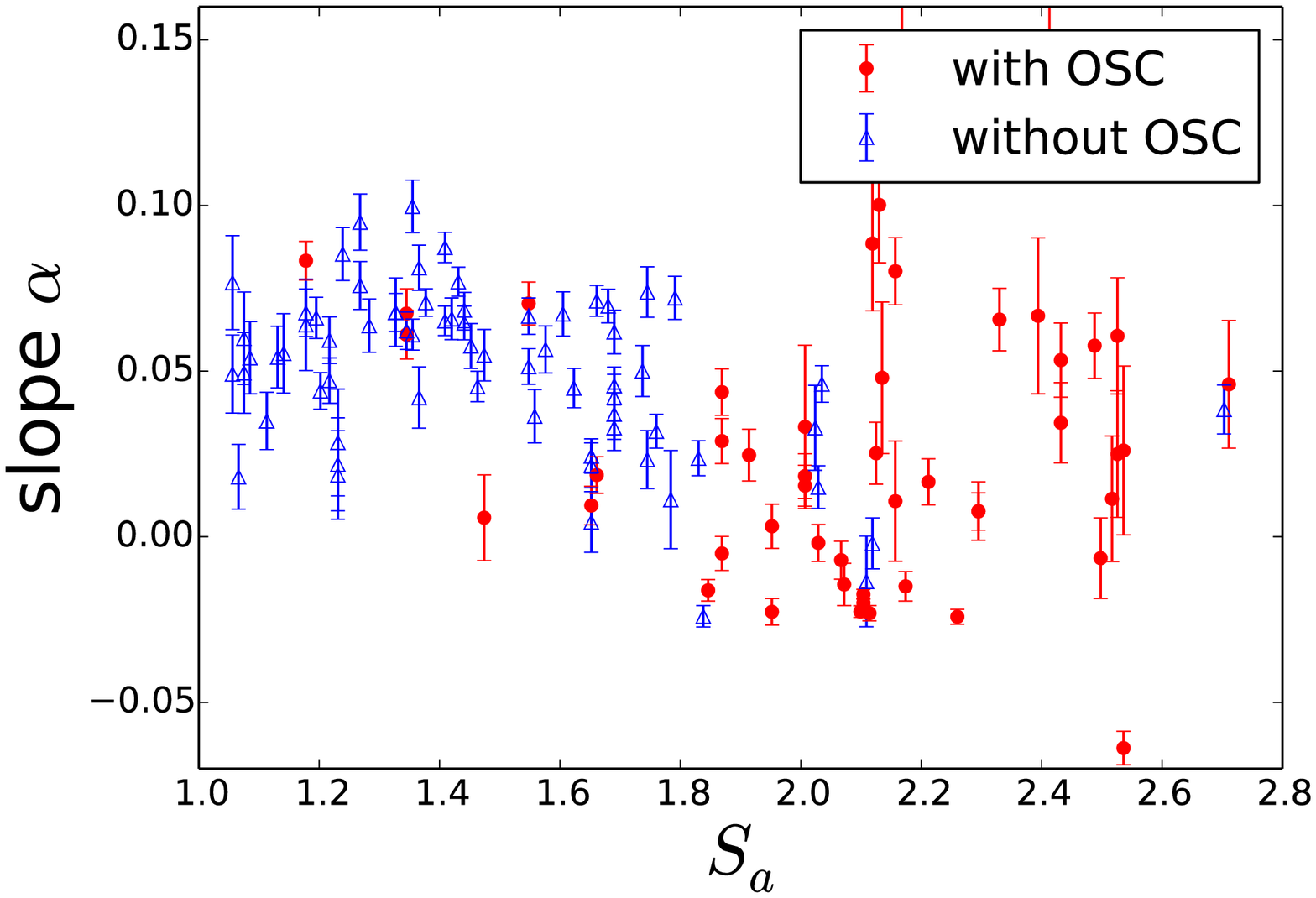}
        \includegraphics[width=3.25in,angle=0]{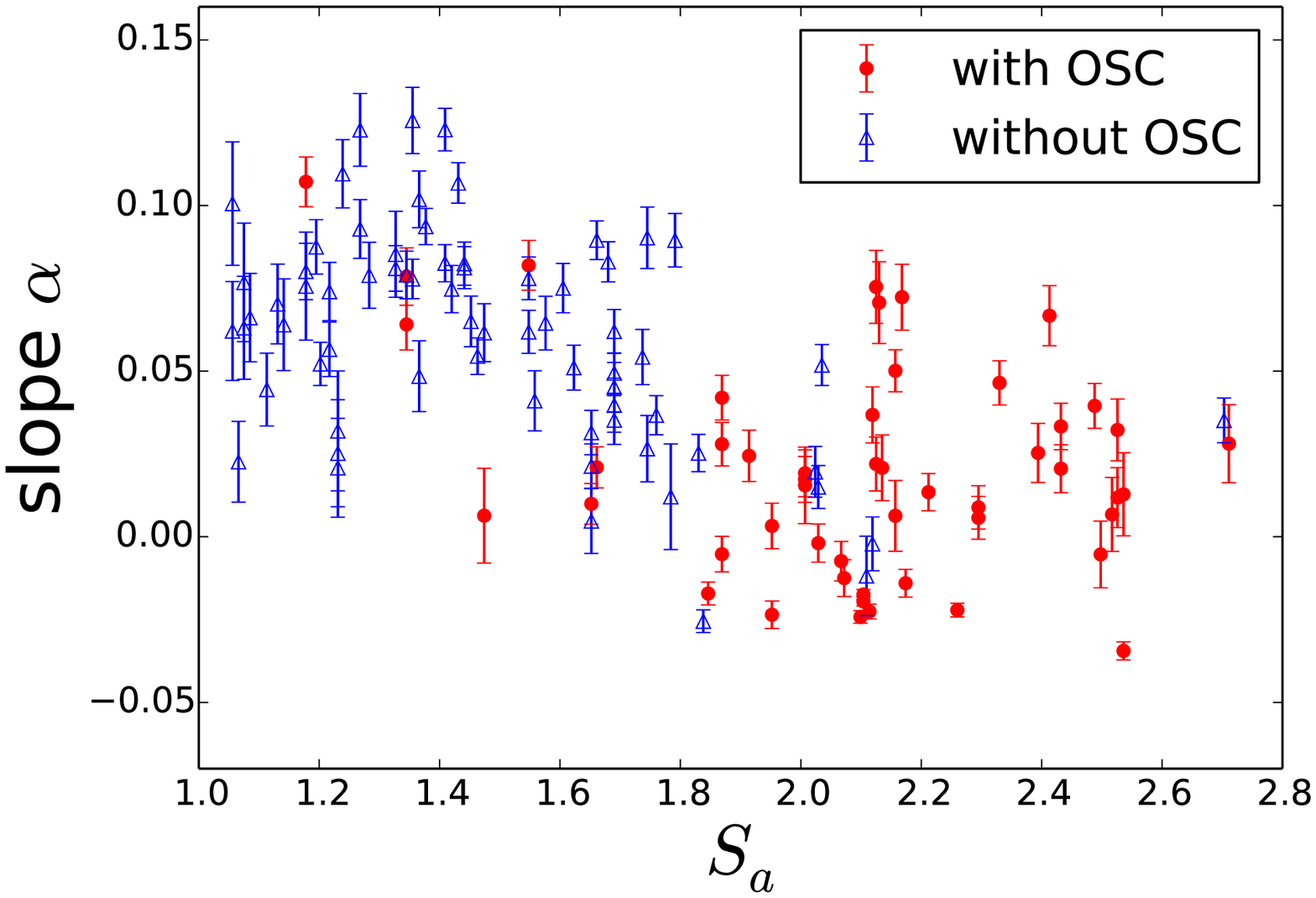}
    \caption{The value of the slope in the $N_{BB}^{-1/4} -$flux plots (Figure \ref{fig_flux_fc1} and 
    \ref{fig_flux_fc2}) as a function of $S_{a}$ for bursts with (red filled circles) and 
without (blue open triangles) oscillations in 4U 1728--34. The slopes are calculated by fitting the 
the points shown in black in Figures \ref{fig_flux_fc1} and \ref{fig_flux_fc2} with a 
linear function in the $N_{BB}^{-1/4} - $flux diagram. In the upper panel the flux is 
normalized by the Eddington flux. In the lower panel the flux is normalized by the peak 
flux in each burst. The error bars represent the  1-$\sigma$ confidence level.
             }
    \label{fig slope vs. sa}
\end{figure}

\begin{figure}
    \centering
        \includegraphics[width=3.20in,angle=0]{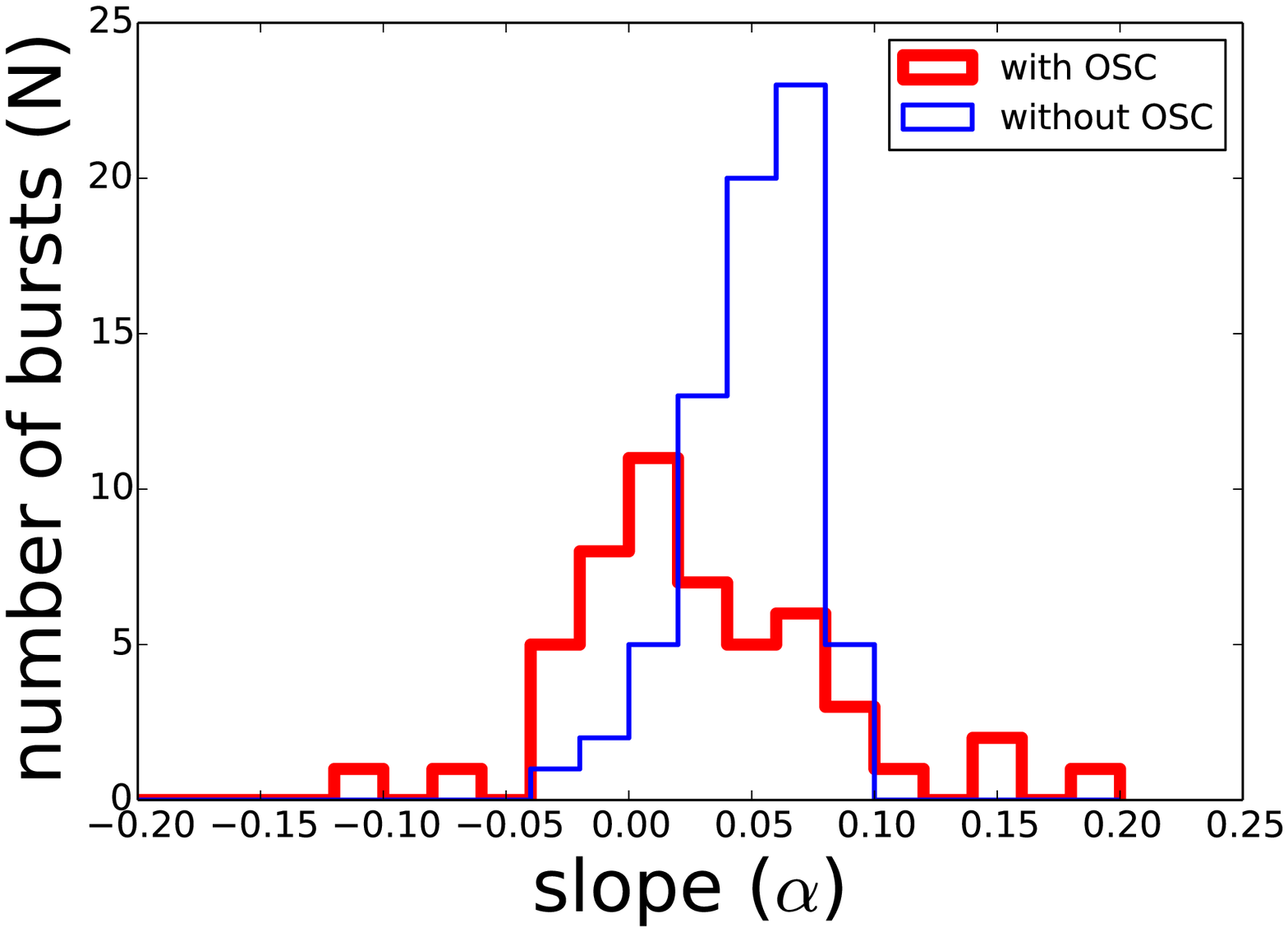}
        \label{fig:hist_sub1}
        \includegraphics[width=3.20in,angle=0]{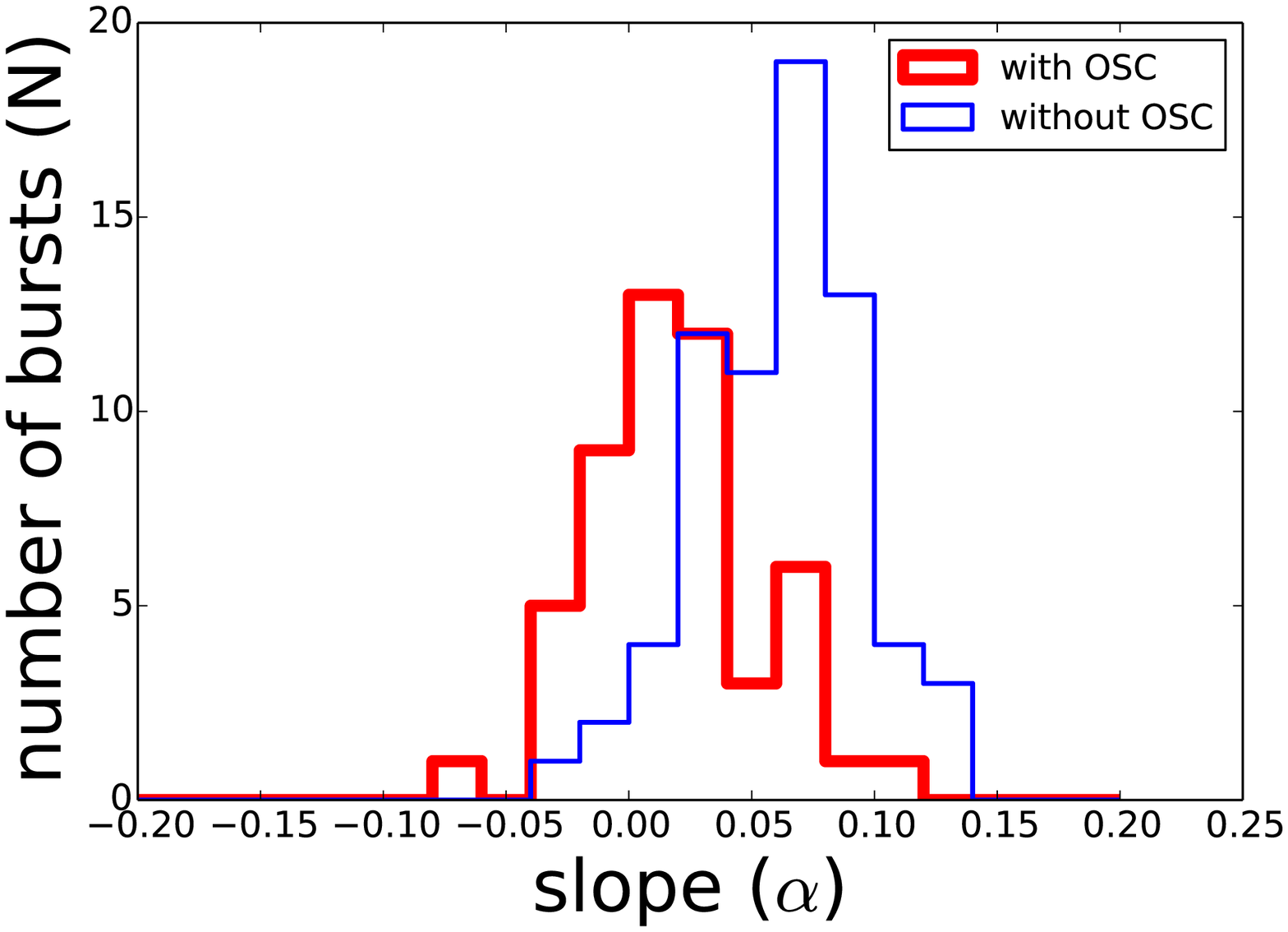}
        \label{fig:hist_sub2}
    \caption{The distribution of slope in the $N_{bb}^{-1/4} -$flux plot 
for bursts with (red) and without (blue) oscillations in 4U 1728--34. In the upper panel 
the flux is normalized by the Eddington flux. In the lower panel the flux is normalized 
by the peak flux in each burst.
             }

    \label{fig hist slope edd}
\end{figure}

In order to quantify this, and to see whether the differences appear 
in all the bursts (including non-PRE) we did the following analysis.
We fitted a linear function to the data during the early cooling phase of the bursts in the 
$N_{BB}^{-1/4} - $flux diagram.  For the PRE bursts, we defined the early cooling 
phase of the bursts as the time interval from the touch down point down to the moment 
when the burst flux returned to 40\% of the touch down flux. For the 
non-PRE bursts we defined the early cooling phase of the bursts as the time interval 
from the peak of the burst to the moment when the burst flux returned to 40\% of 
the peak flux. The early cooling phase as defined in the plots in black filled circles 
in Figure \ref{fig_flux_fc1} and \ref{fig_flux_fc2}

Since different bursts in 4U 1728--34 have slightly different peak or touch-down flux, 
we also used the  Eddington flux to normalize the bursts flux and 
repeated the analysis described above. In this case we used an  Eddington flux of 
$84 \pm 9 \times 10^{-9}$ ergs cm$^{-2}$ s$^{-1}$ \citep[e.g., ][]{Disalvo00, Straaten01, Galloway08a}.

The two  different flux normalizations give slightly different slopes in the fits of the 
$N_{BB}^{-1/4} - $flux plots for each burst. In Figure \ref{fig slope vs. sa} we show 
the fitted slope of each burst as a function of $S_{\rm a}$.  In the upper panel, the 
burst flux is normalized by  the Eddington flux, while in the lower panel the burst 
flux is normalized by the peak flux of each burst. The red filled circles 
indicate bursts without oscillations, while the blue open triangles indicate  bursts
without oscillations. The error bars represent the  1-$\sigma$ confidence level. 

Most of the bursts without oscillations take place at low $S_{\rm a}$ values in the CD, and
have a positive slope in the $N_{BB}^{-1/4} -$flux diagram. On the contrary, most bursts 
with oscillations take place at high values of $S_{\rm a}$, and the slope in the $N_{BB}^{1/4} -$
flux diagram shows both positive and negative values.

\begin{figure*}
    \centering
    \subfigure[convex burst]
    {
        \includegraphics[width=3.30in,angle=0]{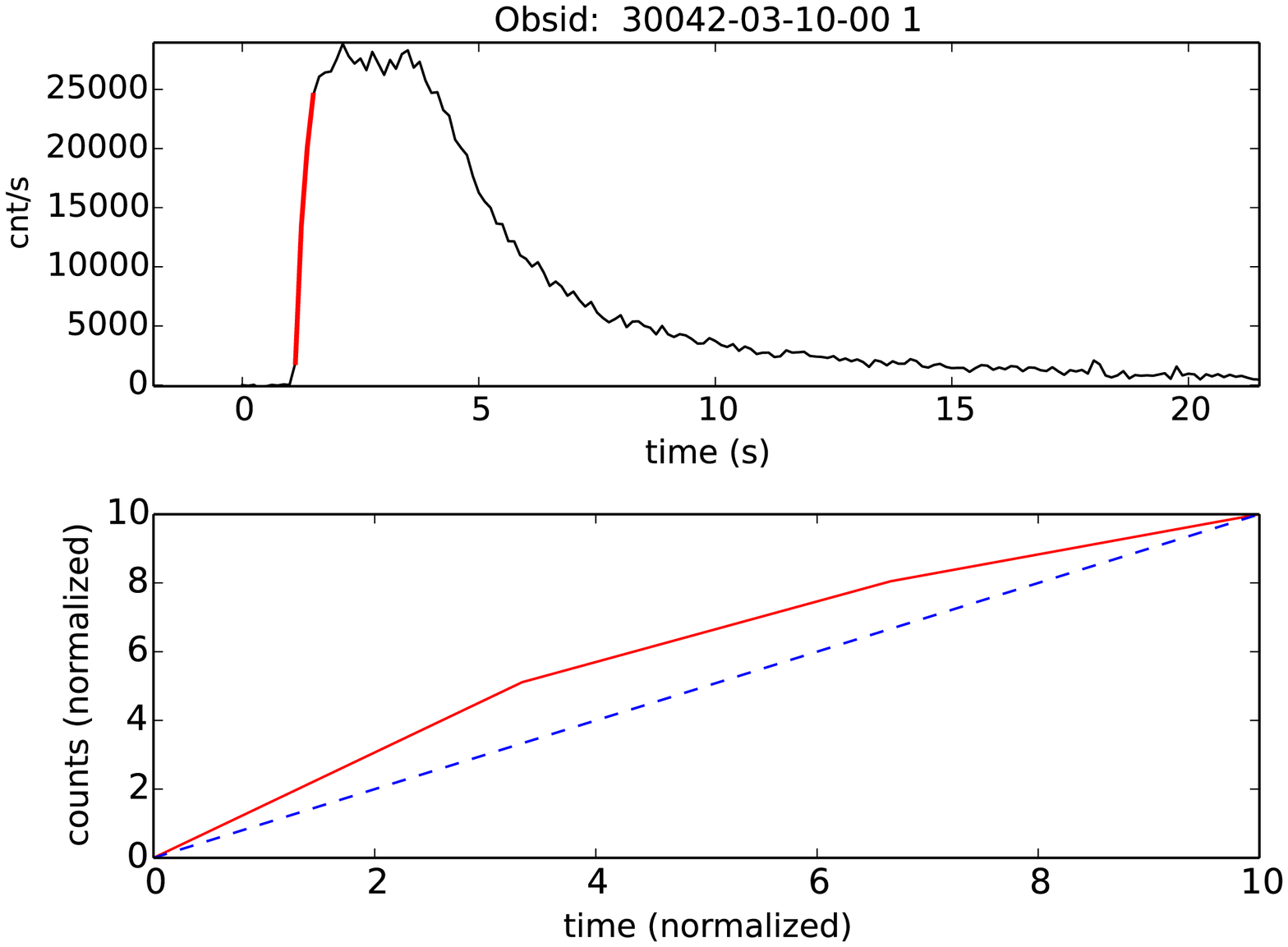}
        \label{conv_gt_0}
    }    
    \subfigure[concave burst]
    {
	\includegraphics[width=3.30in,angle=0]{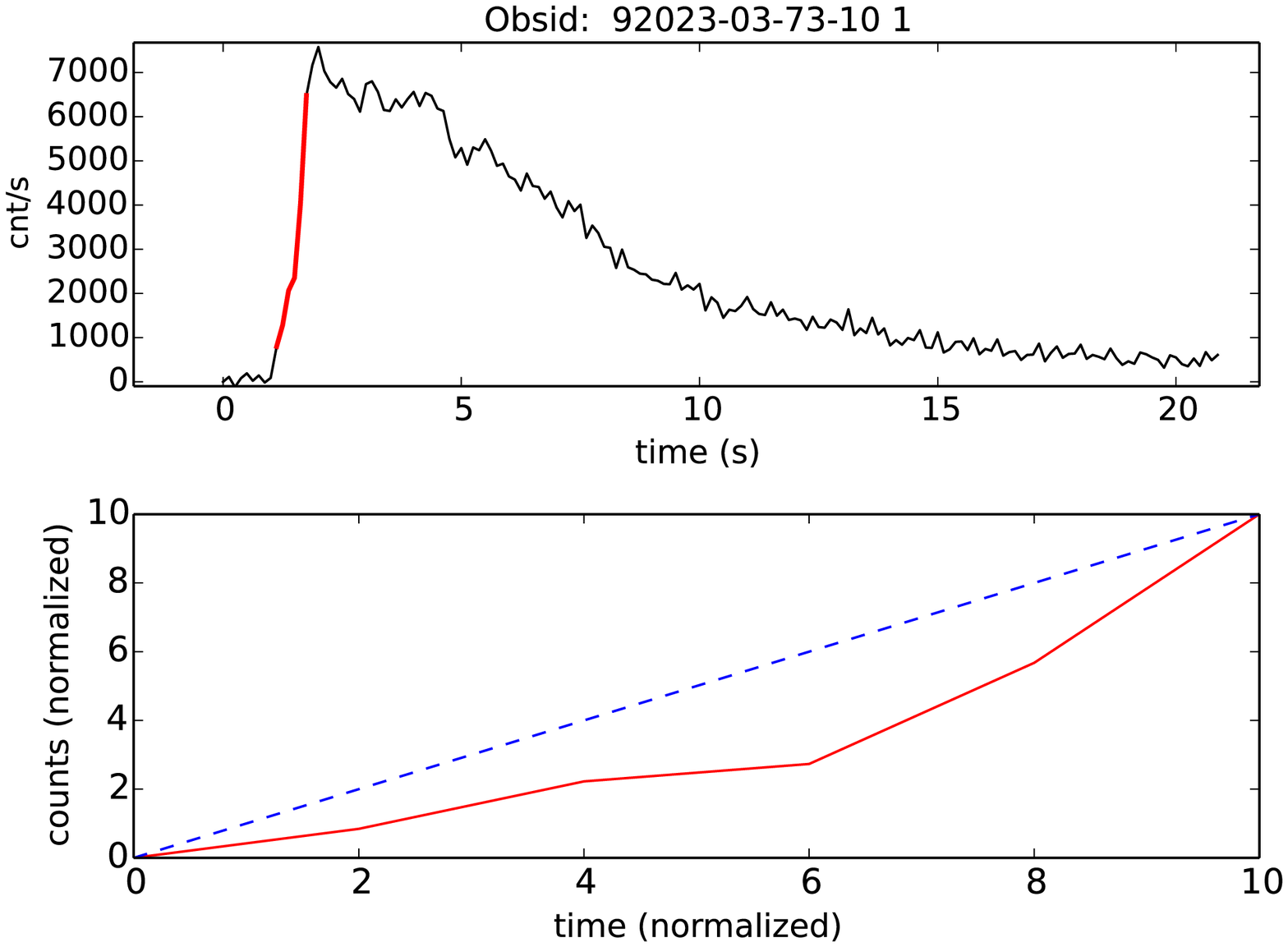}
	\label{conv_lt_0}
    }
    
\caption{Top panel of (a) shows a persistent subtracted burst light curve 
with 0.125s time resolution. The data marked as thick red line are used for convexity 
calculation. The bottom panel of (a) shows the re-normalised light curve
of selected data. The convexity defined in equation \ref{conv} is integrated area 
above or below the diagenal line. The areas above the line is taken to be positive 
and areas below the line negative. Figure (a) shows a convex burst and figure 
(b) shows a concave burst. }

\label{fig conv_plot}
\end{figure*}

Figure \ref{fig hist slope edd} shows the distribution of the slopes of the linear 
fit to the $N_{BB}^{-1/4} -$flux plots, both for bursts with and without oscillations. The burst fluxes
are normalized by the Eddington flux in the upper panel and  by the peak (touch-down) flux in 
the lower panel. It is apparent  in Figure \ref{fig hist slope edd} that 
in bursts without oscillations the slope values are generally above zero, whereas in
bursts with oscillations the slopes are generally around zero. A Kolmogorov--Smirnov (KS) test
gives a probability of  $6.8\times 10^{-5}$ (burst fluxes normalized by Eddington flux) 
and $6.8\times 10^{-8}$ (burst fluxes normalized by the peak flux) that the two samples 
come from the same parent population.

\subsection{X-ray bursts’ light curves}
\label{lightcurve}

\subsubsection{The bursts rise phase--Convexity}

\begin{figure}
    \centering
        \includegraphics[width=3.20in,angle=0]{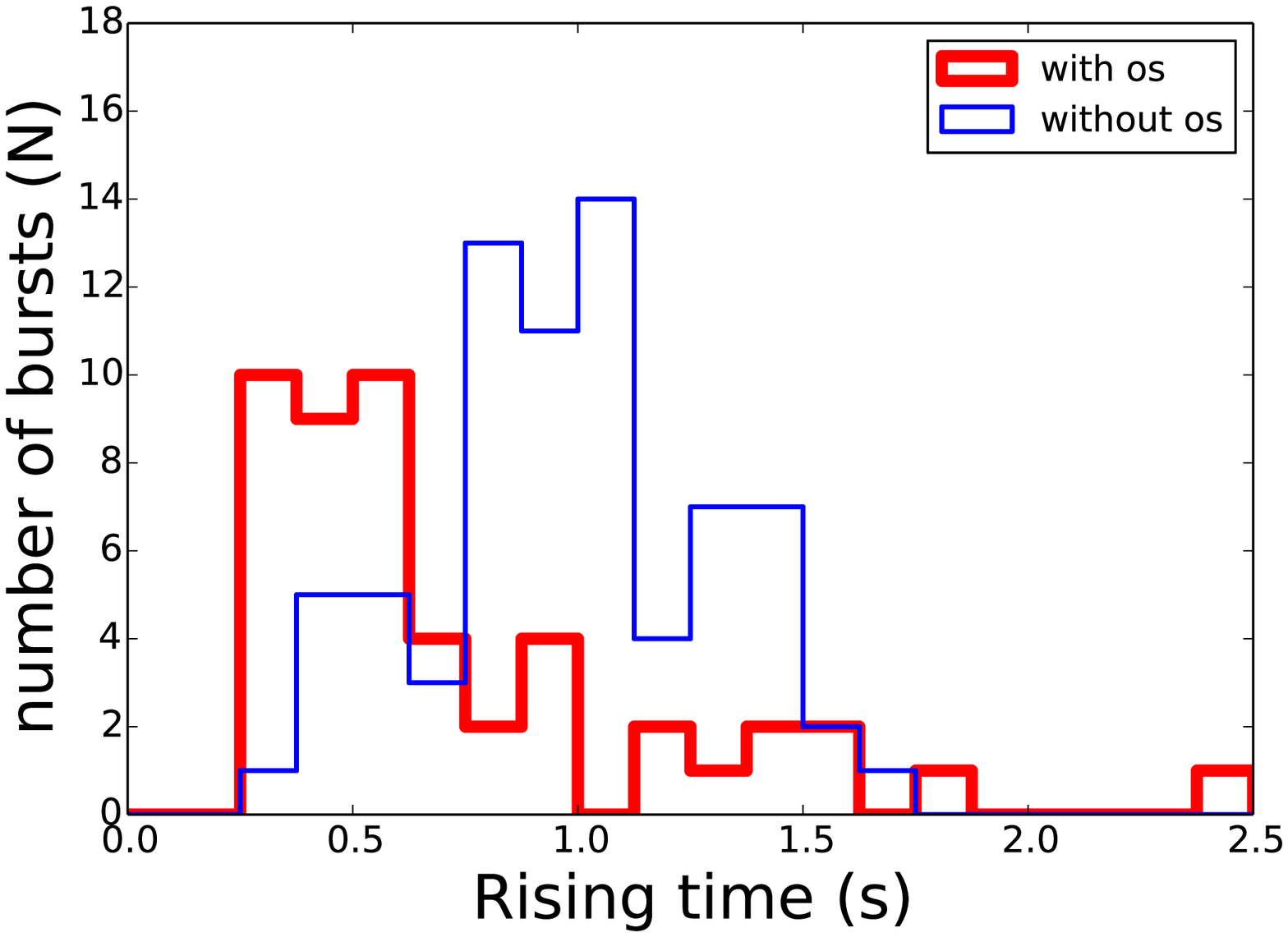}
        \includegraphics[width=3.20in,angle=0]{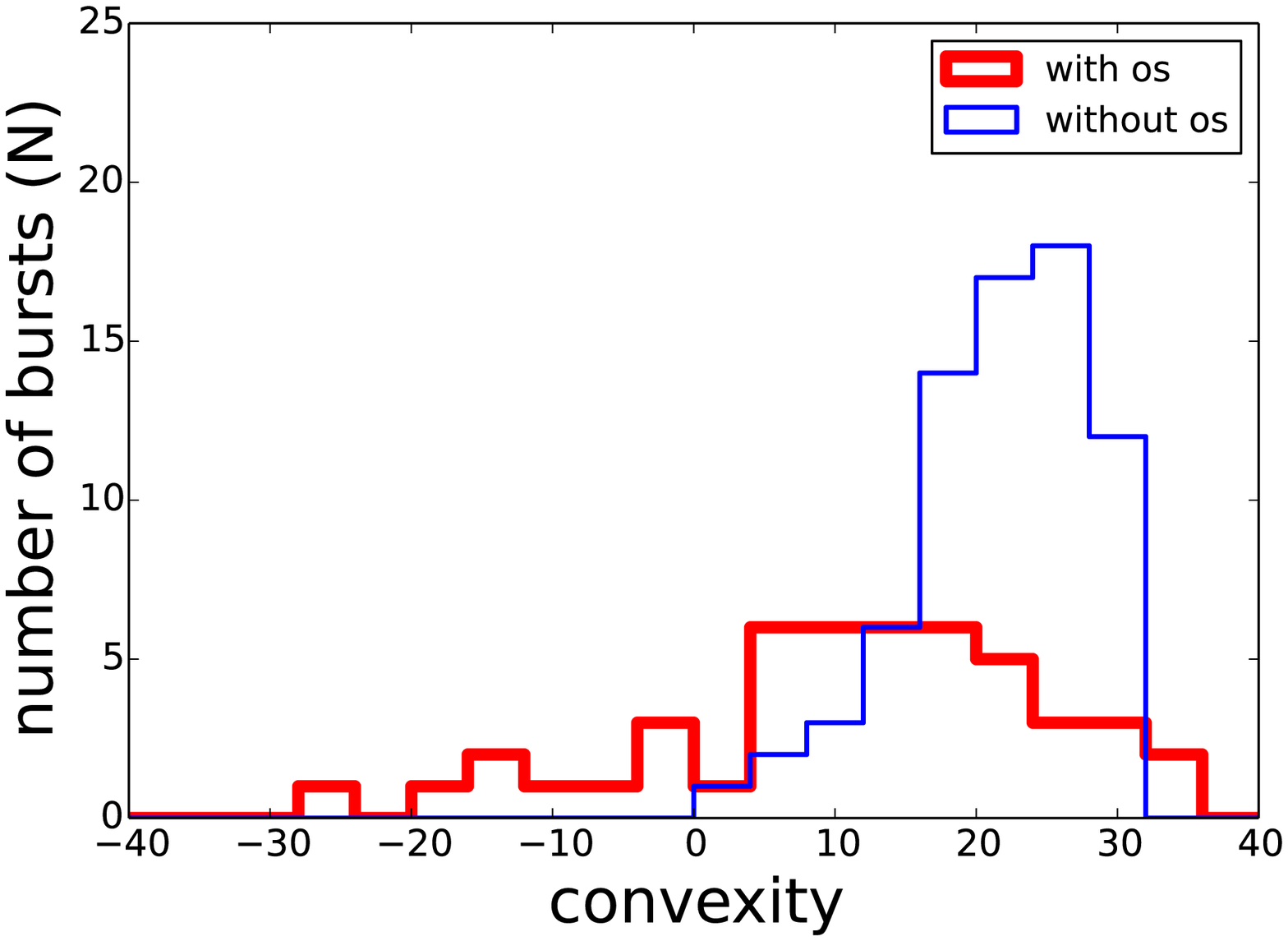}
    \caption{Upper panel: The distribution of the rise time of the bursts with (red) and without
    (blue) burst oscillations in 4U 1728--34. Lower panel: The distribution of the
    convexity in the rise part of the light curve of bursts with and without oscillations in 
    4U 1728--34.
             }

    \label{fig hist convexity}
\end{figure}

We found that the shape of the burst light curve is also different in bursts 
with and without oscillations. In bursts with oscillations 
the flux increases very sharply before it reaches the maximum, and the rise time in 
these bursts is very short. On the contrary, in burst without oscillations the 
flux increases very sharply until about half of the peak flux, and from that point 
onwards the rate of increase of the flux slows down. To quantify this behaviour we used data 
between 10\% and 90\% of the burst  peak flux  to measure the burst rise time. As shown in the 
upper panel of Figure \ref{fig hist convexity}, most of the bursts with oscillations have rise 
times around $\sim 0.5$ s (red histogram), whereas the rise time in bursts without oscillations is 
around $\sim 1.0$ s (blue histogram). The KS probability that the two sample distributions  
come from the same parent population is $6.2 \times 10^{-7}$

In order to describe the shape of the burst rise phase quantitatively, we 
used the convexity ($\mathcal{C}$) parameter in our analysis \citep{Maurer08}. 
{We used the same method  in \citep{Maurer08} to calculate the convexity.  
We used the data where the count rate rises from 10 to 90 percent of the maximum 
count rate (after subtracting the pre-burst emission).  
In order to compare the shape of bursts of different durations and peak
count rates we  normalised the count-rate and time axes so that the burst 
rises from 0 to 10 normalised count-rate units within 10 normalised time units.

\begin{equation}
\mathcal{C} = \sum_{i = 0}^N (c_i - x_i)  \Delta t 
\label{conv}
\end{equation}

In equation \ref{conv},  $c_i$ is the re-normalized count rate in each bin, 
and $x_i$ is the identity function (shown as a diagonal dashed line in the 
lower pannels of (a) and (b) in Figure \ref{fig conv_plot}), 
N is the number of re-normalized time bins and $\Delta t$ is the re-normalized time 
bin size.  Convexity is effectively the integrated area of the
curve above or below the diagonal line - areas above the line being
positive and areas below negative.  The convexity  describes the curvature 
in the light curve rise and quantify whether the curve is convex 
(convexity $> 0$) or concave (convexity $< 0$). 
In Figure \ref{fig conv_plot} we show two examples of bursts with positive and 
negative convexity. 

We  calculated the convexity of the light curves of all X-ray bursts in 4U 1728--34.
In the lower panel of Figure \ref{fig hist convexity} we show the distribution of 
convexity both for bursts with and without oscillations. It is apparent 
that bursts without oscillations always have positive convexity, whereas bursts 
with oscillations show both positive and negative convexity. The KS probability 
that the two sample distributions  come from the 
same parent population is $8.7 \times 10^{-6}$

\subsubsection{The bursts peaking phase}

In the Figure \ref{fig smooth} we show two burst light curves. The light curves
are normalised by the count rate at the peak of the burst. Since it is clear that there are many 
fluctuations during the bursting time, in order to study the bursts profile we smoothed 
these light curves; we show two examples of this in Figure \ref{fig smooth} with red colour. 
We choose three points to characterize the burst profile around the peak: 
Point A indicates the time of the maximum of the burst light curve, and points 
B and C indicate the time at which the light curve of the burst is at 30\% of the 
maximum during the rise and the decay of the burst, respectively. With these
definitions, we call burst duration to the time interval$-$BC,  
rise time to the  time interval$-$BA, and decaying time to the time interval$-$AC.

\begin{figure*}
    \centering
        \subfigure[burst with oscillations]
	{
        \includegraphics[width=3.20in,angle=0]{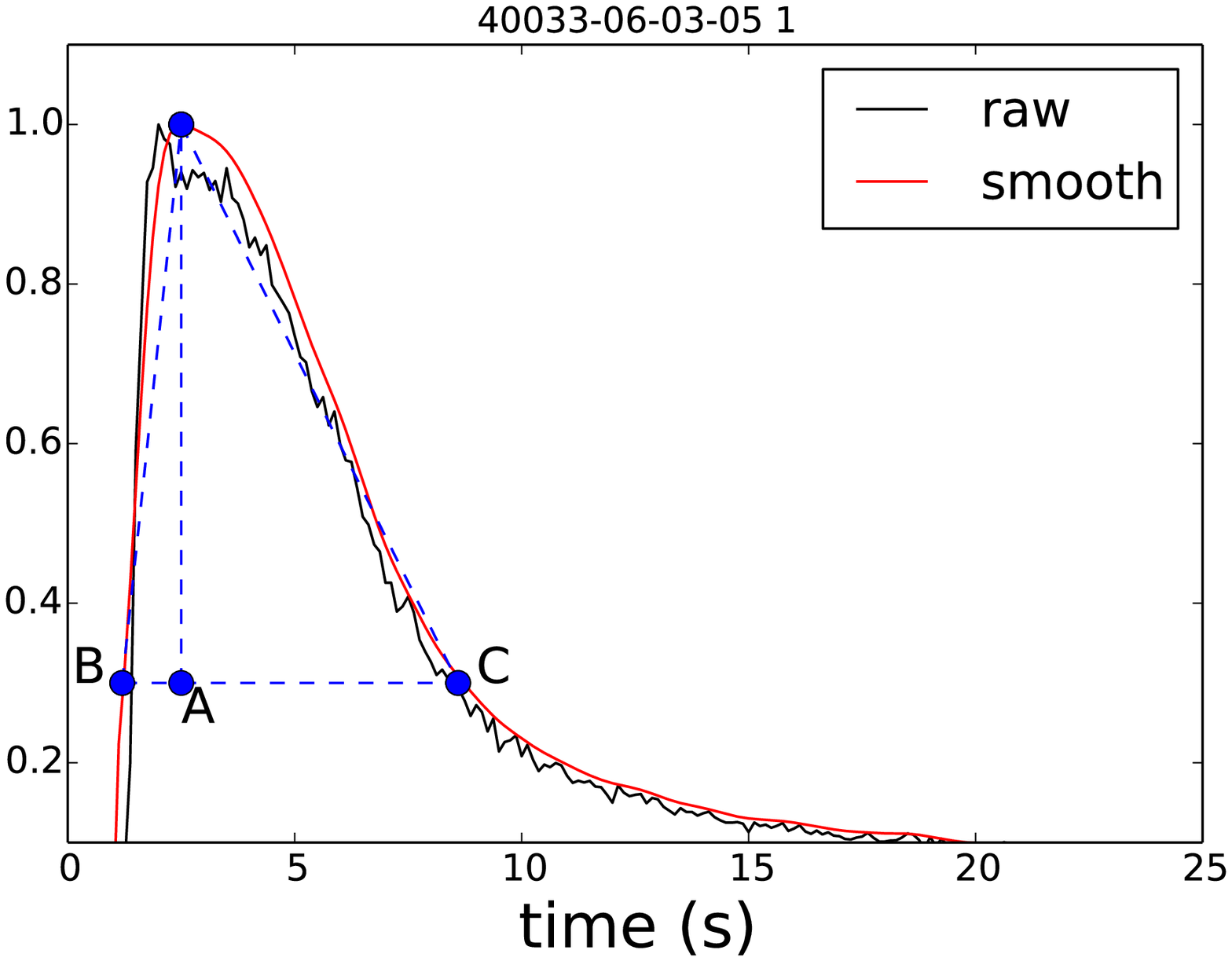}
	}
         \subfigure[burst without oscillations]
	{
        \includegraphics[width=3.20in,angle=0]{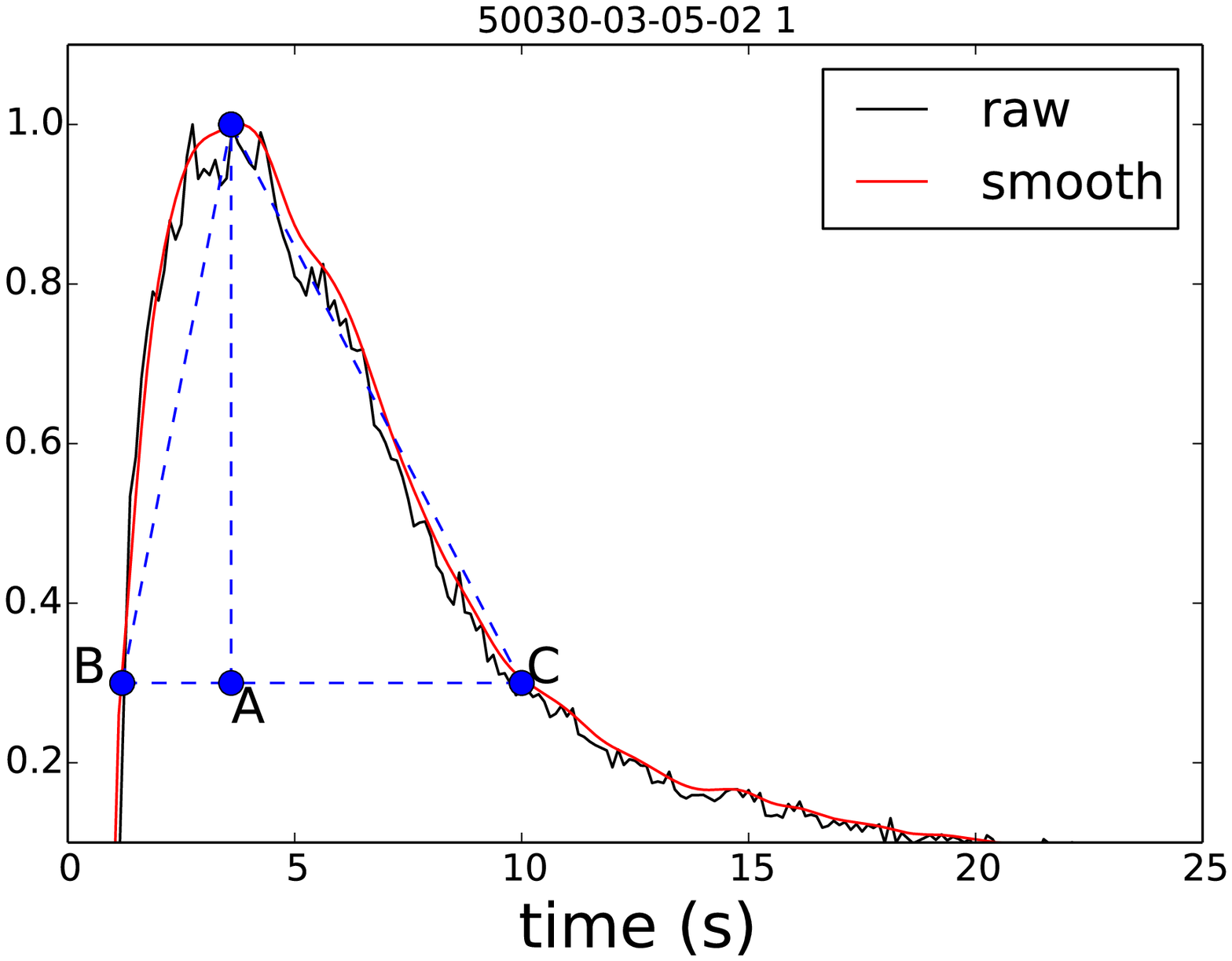}
	}
    \caption{Examples of the light curves of two X-ray bursts in 4U 1728--34. The burst on the left 
    shows burst oscillations whereas the one on the right does not. In both panels we show 
    in black the light curve of the burst normalised to unity at the maximum, and in red 
    the smoothed version of the same light curve.  The point A indicate the time 
    of the maximum of the smoothed light curve, while the points B and C indicate the 
    times at which the flux is 30\% of the maximum during the rising and decaying phases 
    of the burst, respectively.
             }
    \label{fig smooth}
\end{figure*}

In the upper panel of Figure \ref{fig peak_hist} we show the distribution of the burst 
durations. From a KS test (KS probability 0.18) we cannot discard the 
hypothesis that the two types of bursts have the same burst duration. In the lower 
panel of Figure \ref{fig peak_hist} we show the distribution of the rise time 
divided by the burst duration for bursts with (red) and without (blue) oscillations 
in 4U 1728--34. The KS probability that the two samples come from the same parent 
population is $6.4\times10^{-10}$. The distribution of the ratio of rise time to 
the burst duration of bursts with and without oscillations are significantly different.
This indicates that the light-curve profiles of bursts with and without oscillations 
are significantly different,  bursts with oscillations having a more asymmetric 
light-curve profile than bursts without oscillations.

\begin{figure}
    \centering
        \includegraphics[width=3.20in,angle=0]{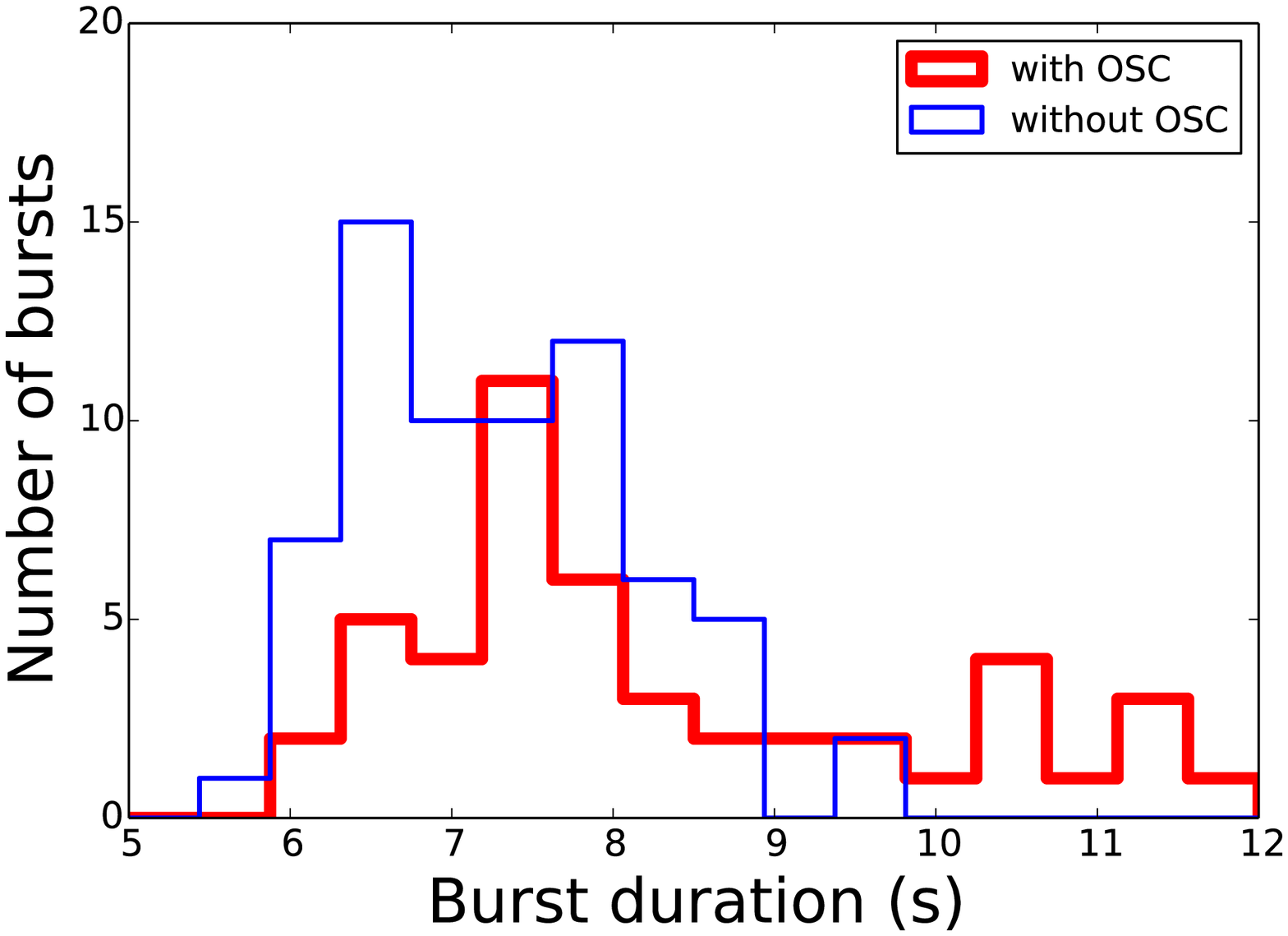}
        \includegraphics[width=3.20in,angle=0]{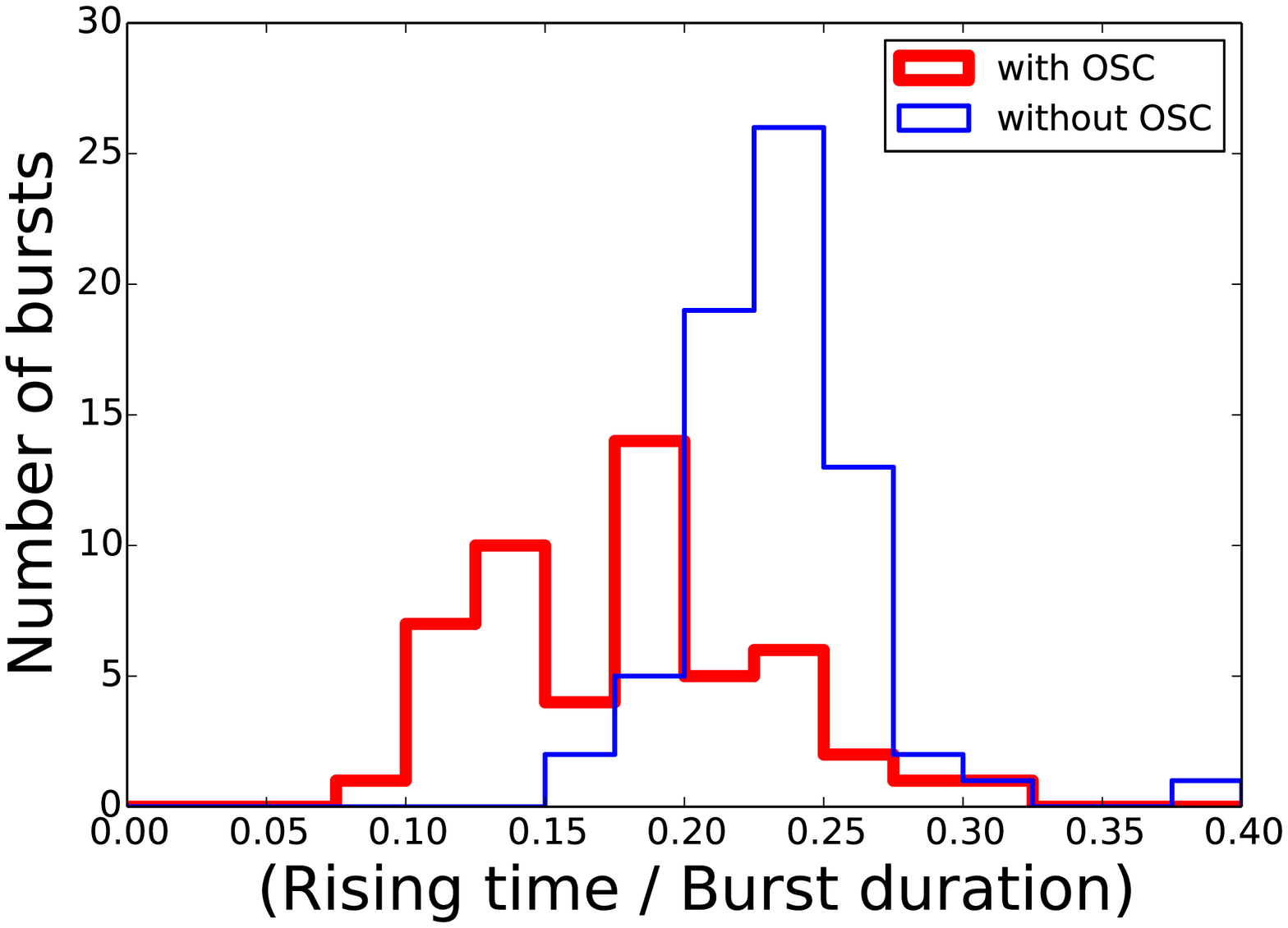}
        \caption{ Upper panel, the distribution of the burst duration, measure as the 
        interval in which the flux is larger than 30\% of the peak flux. Lower panel, 
        the distribution of the ratio of the rise time divided by the burst duration 
        for burst with (red) and without (blue) oscillations.    
             }
    \label{fig peak_hist}
\end{figure}

\section{Discussion}
\label{discussion}

Out of the 121 type-I X-ray bursts from the low-mass X-ray binary 4U 1728$-$34 
observed with RXTE, 73 are photospheric radius expansion (PRE) bursts and 49 show 
burst oscillations at 363 Hz. The convexity of the light curve during the early 
phase of a burst is a good predictor of the presence of oscillations in the tail 
of the burst, regardless of whether the burst shows PRE or not (see lower panel of 
Figure \ref{fig hist convexity}). The distribution of the convexity for bursts 
with and without oscillations is significantly different 
(KS probabilities of $8.7 \times 10^{-6}$). 
Similar to what we found in 4U 1636$-$53 \citep{zhanggb13}, bursts in which the 
blackbody radius remains more or less constant for 2 seconds or more during the 
cooling phase show burst oscillations, whereas those bursts in which the blackbody 
radius changes quickly do not show burst oscillations.

\subsection{Bursts on the colour-colour diagram}
\label{persistent}

Using a sample of 394 bursts in 12 sources, \cite{Muno04} proposed that 
there is a relation between the presence of burst oscillations and accretion rate 
onto the NS. On the other hand, \cite{Zhanggb11} and \cite{Kajava14} studied 
the relation between source accretion state and burst spectral evolution, while 
\cite{zhanggb13} found a link between burst oscillations and burst spectral 
evolution in 4U 1636--53. In this paper we found that this kind of relations are 
much more clear in 4U 1728--34.

Figure \ref{fig ccd} shows the position of the source in the CD at the time when an 
X-ray bursts took place. Bursts with oscillations are located everywhere along the CD, 
but appear preferentially in the so-called lower branch, when the source is in a soft 
state (see also Figure \ref{fig slope vs. sa}). The position of the source along the 
C-like shape in the CD is influenced by the accretion geometry, which itself may 
change with accretion rate and inferred mass accretion rate is high in this area of the CD
\citep[][]{Hasinger89,Mendez99}. This suggests that the appearance of burst oscillations 
is correlated with accretion rate and geometry, however it is not possible to determine 
whether this is a causal relation.

In 4U 1728--34 most of the PRE bursts occur when the inferred mass accretion rate is
low, whereas the PRE bursts in 4U 1636--53 preferentially appear when the 
inferred mass accretion rate is high \citep{Zhanggb11}. Considering the bursts spectral 
properties in 4U 1728--34, 4U 1636--53 and other X-ray bursters \citep{ Straaten01, 
Muno04, Galloway08a, zhanggb13}, the presence of PRE bursts appears to depend upon 
mass accretion rate onto the NS surface, but the effect is different in different sources. 
This is consistent with a difference in the chemical composition of 
the burst fuel in 4U 1636-53 and 4U 1728-34. In sources accreting helium-rich 
material, radius expansion happens at lower accretion rates, because then the 
accumulating fuel layer is colder and builds up to a greater thickness before 
the burst ignites. In sources accreting hydrogen-rich material, radius expansion 
is seen at higher accretion rates when the bursts are helium-like (with short 
durations), whereas it is not seen at lower accretion rates when the bursts 
show evidence of hydrogen (ie. longer durations).

The fact that in 4U 1728--34 bursts with oscillations happen more often 
in the high/soft state, while PRE burst happen mostly in the low/hard state, 
suggests that the oscillations and the PRE phenomenon are both correlated with 
accretion rate and geometry, as proposed by \cite{Franco01, Straaten01, Muno04}; 
however, the presence of bursts with oscillations in the low/hard state in 4U 1728--34 
\citep[and 4U 1636--53; ][]{zhanggb13} indicates that oscillations are not only driven 
by either mass accretion rate or whether the burst shows PRE or not, although 
mass accretion rate (and perhaps PRE) has an influence on the presence of oscillations.

\subsection{X-ray spectral properties and burst oscillations}
\label{specral and oscillations}

We find a strong evidence for a link between burst oscillations and the spectral 
properties during the burst in 4U 1728--34, more specifically between the presence 
of burst oscillations and the changes in the apparent emission area during the decay 
of the burst. The left panels of Figure \ref{fig_spectrum_1} and  \ref{fig_spectrum_2}, 
and the discussion in section \ref{time-resolved spectra}, show that oscillations 
in type-I X-ray bursts in 4U 1728--34 are always associated with a blackbody radius that 
remains more or less constant for at least $\sim$ 2$-$8 s. A similar trend has also been found 
in the LMXB system 4U 1636--53 \citep{zhanggb13}. The changes of the blackbody 
radius could be due to either changes in the apparent emitting area of the NS during 
this phase, or to changes in the colour correction factor, $f_{\rm c}$.

In the case of changes of the apparent emitting area of the NS, \cite{Spitkovsky02} and 
\cite{zhanggb13} suggested that tail oscillations could be due to the spread of a 
cooling wake, which is formed by vortices during the cooling of the NS atmosphere. 
In this scenario, the speed of the cooling wake depends upon the latitude at which 
the burst ignite, since the speed of the cooling front near the equator is higher 
than that near the poles \citep{Spitkovsky02}. 
If the bursts ignite at high latitude on the NS surface, during the burst decay
time the front speed of the cooling wake is slow,   the emission area changes 
slowly, and the asymmetric emission during the burst lasts long.
These bursts would have oscillations and the blackbody radius remains constant 
for a while (see the left panels of Figure  \ref{fig_spectrum_1}). If the bursts 
ignite at low latitude, during the decay time the front speed of the cooling wake 
is fast, the emission area, and hence the blackbody radius,  changes fast, and the 
asymmetric emission is short lived. These bursts would not show oscillations 
and the blackbody radius evolves very fast. We note that this interpretation has not 
been calculated in detail. 

We also note that burst oscillations also appear in the rising phase
of the bursts (e.g., left pannels of Figure \ref{fig_spectrum_2}). The mechanism 
of burst oscillations might be different between the rising and decaying phase. 
The burst oscillations in the rising phase may be caused by asymmetries due to initially
localized nuclear burning that later spreads over the surface of the NS 
\citep[e.g., ][]{Strohmayer96a, Chakraborty14}.

In the case of changes of the $f_{\rm c}$ in the NS atmosphere,  the NS atmosphere model 
of \cite{Suleimanov11} suggests that $f_{\rm c}$ should decrease as the flux decreases
during the burst decay. In section \ref{colour factor} we analysed the $f_{\rm c}$ 
during the burst decay phase assuming a constant emitting NS surface. We found that in 
the early decay phase, the evolution of $f_{\rm c}$ in the bursts without oscillations 
(see the right panels of Figure \ref{fig_flux_fc1} and \ref{fig_flux_fc2}) can be 
qualitatively well-described by the atmosphere models of \cite{Suleimanov11}. 
On the other hand, bursts  with oscillations always show a roughly constant $f_{\rm c}$, 
which disagrees with the atmosphere models.  In particular, these are the hard state 
bursts that \cite{Suleimanov11} have pointed to as being more reliable for 
comparison to spectral models, and thus for use in mass-radius constraints.

In the bursts with oscillations, in order to produce asymmetric emission, only part of 
the NS surface is burning, which contradicts  the assumption in 
\cite{Suleimanov11} that the emission should come from the whole surface of the NS. 
4U 1728--34 is likely a UCXB \citep{Shaposhnikov03, Cumming03, Galloway08a}, so all the 
bursts should ignite in a NS atmosphere with similar chemical composition and have
similar curves in the  $N_{BB}^{-1/4} -$ flux diagram. In Figure \ref{fig hist slope edd}
we show that bursts with a flat $N_{BB}^{-1/4} -$flux relation appear preferentially 
in the soft state.  \cite{Kajava14} found that the X-ray spectrum in the burst decay 
can not be described by the model of \cite{Suleimanov11} when the source is in the soft
state. \cite{Kajava14} suggested that in the soft state the small inner radius of 
the accretion disc and the high mass accretion rate can affect the emission 
from the NS atmosphere. Our results indicate that the behaviour of $f_{\rm c}$ in the 
bursts with oscillations in 4U 1728--34 cannot be explained by the model of \cite{Suleimanov11}.

\subsection{X-ray light curve profiles and burst oscillations}
\label{light curve and oscillation}

The light curve profiles are also different between the bursts with and without oscillations
in 4U 1728--34. In the rise phase of the burst, the light curve of bursts with oscillations 
increases faster than in bursts without oscillations. Around the peak of the burst (e.g. 
within 50\% of the peak flux) the light curve of the bursts without oscillations show smooth 
and symmetric profiles (see right panel of Figure \ref{fig smooth}), whereas, in bursts 
with oscillations the light curves show sharp and asymmetric profiles (see left panel 
of Figure \ref{fig smooth}). The shape of the light curve during the rising phase also shows 
different behaviours in bursts with and without 
oscillations. The convexity takes only positive values in bursts without oscillations 
whereas it shows both positive and negative values in burst with oscillations.

The shape, time scale and peak flux of the light curve of a burst can vary substantially depending
on factors such as the igination latitude, rotation rate of the NS, the accretion rate 
and the composition of the accreted material \citep[e.g.,][]{Taam93, Woosley04, Weinberg06}. 
There are no simple analytic models that take into account all of the relevant parameterst yet. 
\cite{Maurer08} developped a phenomenological model of the burst rise process and showed that
simple measures of the burst morphology can be robust diagnostics of ignition latitude and burning regime.

They simulated burst light curves with ignition at different latitudes on the NS surface, 
and they found that bursts that ignite  near the equator always show positive convexity,
whereas bursts that ignite near the poles show either positive or negative convexity values.
They also found that the chemical composition of the burning material has a
strong influence on convexity, and the rise times increase as the ignition
point moves towards the poles.

4U 1728--34 is an UCXB (H-poor donor),  and has only helium-rich 
bursts with short rise times and similar duration 
\citep{Franco01, Straaten01, Shaposhnikov03, Galloway08a}. Therefore, the chemical 
composition does not have a big influence on the convexity and rise time of the
bursts in 4U 1728--34.  Simulations from \cite{Maurer08} show that 
the different inclinations and source rotation rate have a very small
effect on convexity. So, only the ignition latitude must have a strong influence 
on the burst properties in 4U 1729--34.  This suggests that bursts with oscillations 
in 4U 1728--34 ignite at high  latitudes, whereas bursts without oscillations 
ignite at low latitudes on the NS surface. This is consistent  with the 
findings of \cite{zhanggb13} in the case of 4U 1636--53.

}

\section{conclusion}
\label{conclusion}

We analysed 121 type-I X-ray bursts in the accreting NS LMXB system 4U 1728--34. We found
that:

\begin{itemize}

\item The rise time and convexity in bursts with and without oscillations are different.
Bursts with oscillations have short rise time and both positive and negative 
convexity values, whereas bursts without oscillations have long rise time and only positive 
convexity values.

\item  Around the peak of the burst, the light curves are more asymmetric in bursts with
oscillations than in bursts without oscillations.

\item In the early decay phase of the bursts, the energy spectrum is different in 
bursts with and without oscillations. The spectrum can be explained by the 
NS atmosphere model only in the bursts without oscillations.

\item During the cooling phase, bursts in which the blackbody radius remains more or 
less constant for $> 2$ s show coherent oscillations, whereas bursts in which the 
blackbody radius changes rapidly show no coherent oscillations.

\item  Bursts with oscillations ignite at high latitude when mass accretion rate onto 
the NS surface is high, whereas burst without oscillations ignite at low latitude
when mass accretion rate is low.

\end{itemize}

\section{Appendix}

\begin{table*}
\begin{center}
\caption{All parameters of type-I X-ray bursts in 4U 1728--34 observed with RXTE. 
The columns in the Table are: RXTE OBSID, MJD,
soft colour (SC), hard colour (HC), intensity (IT), 
whether the burst is PRE or not, 
whether the burst had oscillation, peak flux ($10^{-8}$ ergs cm$^{-2}$ s$^{-1}$),  
S$_{a}$ value at the time of the onset of the burst, 
rise time (s), convexity.  }
\begin{tabular}{ccccccccccccc}

\hline \hline
obsid & MJD & SC & HC & IT  & PRE  & OS & peak flux & S$_{a}$ & T$_{rise}$ & convexity \\
\hline

  10073-01-01-000  & 50128.74881663  & 1.444 & 0.842 & 0.133 &   1 &   1 & 7.00 & 1.90 & 0.500 & 17.24  \\
   10073-01-01-00  & 50128.88223239  & 1.439 & 0.826 & 0.137 &   1 &   1 & 4.33 & 1.95 & 0.375 & 7.60  \\
  10073-01-02-000  & 50129.16474396  & 1.446 & 0.801 & 0.147 &   0 &   0 & 6.38 & 2.02 & 1.375 & 14.33  \\
   10073-01-02-00  & 50129.28569304  & 1.441 & 0.778 & 0.158 &   0 &   1 & 6.62 & 2.10 & 0.375 & 8.27  \\
   10073-01-02-00  & 50129.41724396  & 1.441 & 0.778 & 0.158 &   0 &   1 & 6.59 & 2.10 & 0.250 & 5.64  \\
  10073-01-03-000  & 50129.81060623  & 1.435 & 0.779 & 0.156 &   0 &   1 & 6.32 & 2.10 & 0.625 & 18.34  \\
  10073-01-04-000  & 50131.73050785  & 1.421 & 0.861 & 0.112 &   0 &   1 & 6.55 & 1.87 & 0.625 & 34.97  \\
   10073-01-04-00  & 50131.89509119  & 1.451 & 0.895 & 0.103 &   0 &   0 & 6.15 & 1.76 & 0.750 & 25.50  \\
   10073-01-06-00  & 50135.96476711  & 1.581 & 0.978 & 0.073 &   1 &   0 & 5.61 & 1.43 & 1.250 & 30.74  \\
   10073-01-07-00  & 50137.24061202  & 1.592 & 0.982 & 0.076 &   1 &   0 & 5.52 & 1.41 & 1.375 & 31.08  \\
  10073-01-08-000  & 50137.74437359  & 1.598 & 0.994 & 0.079 &   1 &   0 & 5.45 & 1.38 & 1.250 & 22.99  \\
  10073-01-09-000  & 50138.97027058  & 1.615 & 0.994 & 0.084 &   1 &   0 & 5.59 & 1.35 & 0.875 & 27.87  \\
   20083-01-01-01  & 50710.52293840  & 1.514 & 0.725 & 0.160 &   0 &   1 & 7.24 & 2.49 & 1.500 & 32.40  \\
  20083-01-01-020  & 50711.42286896  & 1.472 & 0.727 & 0.148 &   0 &   1 & 6.67 & 2.41 & 1.250 & 5.11  \\
   20083-01-02-01  & 50712.65664789  & 1.476 & 0.732 & 0.156 &   0 &   1 & 7.21 & 2.43 & 0.500 & 15.43  \\
   20083-01-02-01  & 50712.75776479  & 1.476 & 0.732 & 0.156 &   0 &   1 & 6.73 & 2.43 & 0.625 & 10.93  \\
  20083-01-02-000  & 50713.27982498  & 1.462 & 0.737 & 0.146 &   0 &   1 & 6.83 & 2.39 & 1.375 & 22.17  \\
   20083-01-04-00  & 50717.61406688  & 1.378 & 0.753 & 0.129 &   0 &   1 & 7.44 & 2.16 & 0.875 & 19.16  \\
   20083-01-04-00  & 50717.72907845  & 1.378 & 0.753 & 0.129 &   0 &   1 & 6.32 & 2.16 & 0.250 & 10.74  \\
   20083-01-04-01  & 50718.47167104  & 1.379 & 0.814 & 0.106 &   1 &   1 & 7.18 & 2.01 & 0.375 & 20.93  \\
   20083-01-04-01  & 50718.66260275  & 1.379 & 0.814 & 0.106 &   1 &   1 & 7.02 & 2.01 & 0.375 & 22.41  \\
   30042-03-01-00  & 51086.42292104  & 1.721 & 1.058 & 0.133 &   1 &   0 & 5.70 & 1.08 & 0.875 & 25.72  \\
   30042-03-03-01  & 51110.10594766  & 1.689 & 1.077 & 0.117 &   1 &   0 & 5.64 & 1.07 & 1.000 & 24.71  \\
   30042-03-06-00  & 51118.15833771  & 1.704 & 1.076 & 0.125 &   1 &   0 & 5.76 & 1.06 & 1.500 & 26.25  \\
   30042-03-06-00  & 51118.29117336  & 1.704 & 1.076 & 0.125 &   1 &   0 & 6.06 & 1.06 & 1.250 & 23.04  \\
   30042-03-07-01  & 51119.94659002  & 1.697 & 1.072 & 0.125 &   1 &   0 & 5.90 & 1.07 & 1.000 & 10.57  \\
   30042-03-07-00  & 51120.08506803  & 1.696 & 1.080 & 0.124 &   1 &   0 & 5.93 & 1.07 & 1.500 & 28.06  \\
   30042-03-10-00  & 51127.81510275  & 1.707 & 1.043 & 0.133 &   1 &   0 & 5.88 & 1.14 & 0.500 & 13.40  \\
   30042-03-11-00  & 51128.02614442  & 1.693 & 1.034 & 0.133 &   1 &   0 & 5.60 & 1.18 & 1.000 & 26.63  \\
   30042-03-12-00  & 51128.68105183  & 1.674 & 1.021 & 0.134 &   1 &   0 & 5.90 & 1.23 & 0.500 & 16.79  \\
   30042-03-12-00  & 51128.81400322  & 1.674 & 1.021 & 0.134 &   1 &   0 & 6.12 & 1.23 & 0.750 & 20.02  \\
   30042-03-13-00  & 51129.01389326  & 1.672 & 1.022 & 0.134 &   0 &   0 & 6.42 & 1.23 & 0.875 & 29.16  \\
   30042-03-14-00  & 51133.42397428  & 1.453 & 0.859 & 0.164 &   1 &   0 & 6.71 & 1.84 & 0.250 & 0.00  \\
   30042-03-15-00  & 51133.67302521  & 1.457 & 0.883 & 0.155 &   0 &   0 & 6.31 & 1.78 & 0.375 & 17.85  \\
   30042-03-17-00  & 51134.57236549  & 1.556 & 0.975 & 0.124 &   1 &   1 & 6.49 & 1.47 & 0.250 & 8.79  \\
   30042-03-20-00  & 51196.99073354  & 1.670 & 1.063 & 0.100 &   1 &   0 & 5.82 & 1.13 & 1.375 & 17.00  \\
   40033-06-01-00  & 51198.14346502  & 1.679 & 1.068 & 0.105 &   1 &   0 & 5.49 & 1.11 & 1.000 & 21.20  \\
   40033-06-02-00  & 51200.26718609  & 1.677 & 1.031 & 0.115 &   1 &   0 & 6.09 & 1.20 & 0.875 & 24.86  \\
   40033-06-02-01  & 51201.99242914  & 1.629 & 0.995 & 0.123 &   1 &   1 & 6.23 & 1.34 & 0.875 & 26.11  \\
   40033-06-02-03  & 51204.00120229  & 1.509 & 0.922 & 0.138 &   1 &   0 & 3.68 & 1.65 & 0.750 & 27.00  \\
   40033-06-02-03  & 51204.12994072  & 1.509 & 0.922 & 0.138 &   1 &   1 & 6.96 & 1.65 & 0.375 & 16.00  \\
   40033-06-02-04  & 51205.98997544  & 1.497 & 0.923 & 0.130 &   1 &   0 & 6.50 & 1.65 & 0.375 & 14.14  \\
   40033-06-02-05  & 51206.14071618  & 1.487 & 0.926 & 0.129 &   1 &   1 & 7.08 & 1.66 & 0.250 & 8.48  \\
   40033-06-03-01  & 51208.98528215  & 1.494 & 0.926 & 0.116 &   1 &   0 & 7.75 & 1.65 & 0.750 & 21.48  \\
  40033-06-03-020  & 51209.91810044  & 1.442 & 0.853 & 0.131 &   1 &   1 & 6.83 & 1.87 & 0.250 & 4.02  \\
  40033-06-03-020  & 51210.08248701  & 1.442 & 0.853 & 0.131 &   1 &   1 & 6.94 & 1.87 & 0.250 & 14.21  \\
   40033-06-03-05  & 51213.93852289  & 1.436 & 0.823 & 0.124 &   1 &   1 & 7.19 & 1.95 & 0.250 & 10.13  \\
   40027-06-01-00  & 51236.79212243  & 1.536 & 0.952 & 0.077 &   1 &   0 & 6.00 & 1.55 & 1.000 & 22.08  \\
   40027-06-01-02  & 51237.20253331  & 1.536 & 0.950 & 0.078 &   0 &   0 & 6.53 & 1.56 & 1.125 & 30.75  \\
   40027-06-01-03  & 51238.79158424  & 1.570 & 0.980 & 0.079 &   1 &   0 & 5.90 & 1.44 & 1.375 & 23.95  \\
   40027-06-01-08  & 51240.04664211  & 1.592 & 0.977 & 0.083 &   1 &   0 & 5.91 & 1.42 & 0.750 & 19.92  \\
   40027-08-01-01  & 51359.82659581  & 1.475 & 0.946 & 0.085 &   1 &   0 & 6.42 & 1.62 & 1.250 & 30.84  \\
   40027-08-03-00  & 51369.42190252  & 1.582 & 1.016 & 0.081 &   1 &   0 & 5.66 & 1.34 & 1.000 & 13.63  \\
  40019-03-02-000  & 51409.39851711  & 1.514 & 0.739 & 0.242 &   0 &   1 & 6.51 & 2.53 & 2.375 & 27.45  \\
  40019-03-02-000  & 51409.50654951  & 1.514 & 0.739 & 0.242 &   0 &   1 & 6.02 & 2.53 & 0.500 & -16.71  \\
   40019-03-02-00  & 51409.58388748  & 1.533 & 0.734 & 0.250 &   0 &   1 & 6.39 & 2.54 & 0.375 & 5.54  \\
   40019-03-02-00  & 51409.65764905  & 1.533 & 0.734 & 0.250 &   0 &   1 & 5.71 & 2.54 & 0.500 & -27.08  \\
   40019-03-01-06  & 51443.21822775  & 1.449 & 0.867 & 0.097 &   0 &   0 & 6.95 & 1.83 & 0.375 & 9.25  \\
   50023-01-07-00  & 51628.86897428  & 1.496 & 0.814 & 0.123 &   0 &   0 & 5.92 & 2.70 & 0.750 & 12.11  \\

\hline 
\end{tabular}
\label{tab:pileup}
\begin{tablenotes}
\item[]
\end{tablenotes}
\end{center}
\end{table*}

\begin{table*}{-1}
\begin{center}
\caption{Continued }
\begin{tabular}{ccccccccccccc}

\hline \hline
obsid & MJD & SC & HC & IT  & PRE  & OS & peak flux & S$_{a}$ & T$_{rise}$ & convexity \\
\hline
   50029-23-02-01  & 51657.20330299  & 1.553 & 0.987 & 0.070 &   1 &   0 & 6.16 & 1.44 & 1.250 & 21.60  \\
   50029-23-02-02  & 51657.67851132  & 1.555 & 0.978 & 0.072 &   0 &   0 & 6.69 & 1.46 & 1.000 & 21.89  \\
   50023-01-21-00  & 51691.71255646  & 1.605 & 1.011 & 0.079 &   1 &   0 & 5.87 & 1.33 & 1.125 & 25.10  \\
   50023-01-22-00  & 51695.33975553  & 1.596 & 0.998 & 0.084 &   1 &   0 & 5.62 & 1.37 & 1.375 & 24.12  \\
   50023-01-23-00  & 51697.47930414  & 1.588 & 1.006 & 0.085 &   1 &   0 & 5.54 & 1.35 & 1.000 & 22.84  \\
  50030-03-02-000  & 51942.94612127  & 1.380 & 0.802 & 0.060 &   1 &   0 & 6.26 & 2.04 & 0.500 & 23.77  \\
   50030-03-03-02  & 51949.12603447  & 1.525 & 1.024 & 0.050 &   1 &   0 & 5.66 & 1.41 & 1.375 & 29.22  \\
   50030-03-04-00  & 52007.61316410  & 1.433 & 0.866 & 0.124 &   1 &   1 & 6.91 & 1.85 & 1.125 & 28.34  \\
   50030-03-04-02  & 52008.08734234  & 1.394 & 0.803 & 0.142 &   1 &   1 & 7.08 & 2.03 & 0.500 & 15.23  \\
   50030-03-05-03  & 52024.43797313  & 1.626 & 1.059 & 0.079 &   1 &   0 & 5.69 & 1.20 & 1.125 & 17.63  \\
   50030-03-05-02  & 52024.69547313  & 1.643 & 1.057 & 0.079 &   1 &   0 & 5.27 & 1.18 & 1.375 & 23.81  \\
   60029-02-01-00  & 52056.40803678  & 1.516 & 0.860 & 0.137 &   1 &   0 & 5.95 & 1.79 & 1.125 & 18.50  \\
   50030-03-06-00  & 52112.25357498  & 1.481 & 0.927 & 0.050 &   1 &   0 & 5.51 & 1.66 & 1.625 & 24.46  \\
   50030-03-06-02  & 52112.58463400  & 1.475 & 0.920 & 0.052 &   0 &   0 & 6.01 & 1.68 & 1.000 & 16.85  \\
   50030-03-08-02  & 52200.14829720  & 1.392 & 0.749 & 0.087 &   0 &   1 & 6.86 & 2.17 & 0.250 & 1.45  \\
   50030-03-08-00  & 52200.39319882  & 1.413 & 0.743 & 0.090 &   0 &   1 & 6.82 & 2.21 & 0.500 & 15.00  \\
   50030-03-09-01  & 52209.99568725  & 1.441 & 0.713 & 0.091 &   0 &   1 & 7.61 & 2.33 & 1.500 & 18.86  \\
  70028-01-01-070  & 52336.21127174  & 1.473 & 0.919 & 0.093 &   1 &   0 & 6.74 & 1.69 & 0.875 & 22.83  \\
  70028-01-01-070  & 52336.35301942  & 1.473 & 0.919 & 0.093 &   0 &   0 & 6.39 & 1.69 & 0.875 & 16.87  \\
  70028-01-01-070  & 52336.49959928  & 1.473 & 0.919 & 0.093 &   0 &   0 & 6.40 & 1.69 & 0.875 & 27.85  \\
   70028-01-01-02  & 52337.09745229  & 1.481 & 0.914 & 0.090 &   1 &   0 & 6.55 & 1.69 & 0.750 & 29.73  \\
   70028-01-01-00  & 52337.94569303  & 1.464 & 0.899 & 0.090 &   1 &   0 & 6.35 & 1.74 & 0.750 & 30.13  \\
   70028-01-01-10  & 52338.26090137  & 1.487 & 0.913 & 0.087 &   0 &   0 & 6.48 & 1.69 & 0.625 & 12.28  \\
   70028-01-01-12  & 52338.41596502  & 1.501 & 0.943 & 0.084 &   1 &   0 & 6.08 & 1.60 & 0.625 & 26.34  \\
   90406-01-01-00  & 53076.07050785  & 1.376 & 0.761 & 0.212 &   0 &   1 & 6.14 & 2.13 & 0.500 & -1.54  \\
   91023-01-02-00  & 53465.26385854  & 1.364 & 0.763 & 0.189 &   0 &   1 & 7.13 & 2.12 & 0.375 & -0.75  \\
   92023-03-03-00  & 53802.03840137  & 1.492 & 0.959 & 0.097 &   1 &   0 & 6.28 & 1.58 & 0.750 & 25.20  \\
   92023-03-23-00  & 53842.51789211  & 1.372 & 0.766 & 0.142 &   0 &   1 & 6.43 & 2.12 & 0.625 & -8.40  \\
   92023-03-35-00  & 53866.02735970  & 1.488 & 0.973 & 0.096 &   1 &   0 & 5.87 & 1.55 & 0.750 & 24.25  \\
   92023-03-41-00  & 53878.55446039  & 1.372 & 0.769 & 0.176 &   1 &   1 & 6.85 & 2.11 & 0.250 & 7.33  \\
   92023-03-52-00  & 53900.74835507  & 1.427 & 0.735 & 0.163 &   0 &   1 & 6.72 & 2.26 & 1.125 & 30.51  \\
   92023-03-76-00  & 53948.69922313  & 1.379 & 0.786 & 0.188 &   0 &   1 & 6.99 & 2.07 & 1.750 & 22.11  \\
   92023-03-83-00  & 53962.32031109  & 1.360 & 0.741 & 0.145 &   0 &   1 & 6.48 & 2.17 & 0.375 & -6.70  \\
   92023-03-87-01  & 53970.37473817  & 1.377 & 0.789 & 0.138 &   0 &   1 & 7.24 & 2.07 & 0.875 & 13.34  \\
   92023-03-95-00  & 53986.15183308  & 1.421 & 0.807 & 0.183 &   0 &   1 & 6.55 & 2.01 & 0.875 & -2.07  \\
   92023-03-02-10  & 53996.50301363  & 1.576 & 1.044 & 0.099 &   1 &   0 & 6.13 & 1.28 & 1.000 & 25.42  \\
   92023-03-06-10  & 54004.54995808  & 1.447 & 0.903 & 0.186 &   1 &   0 & 7.10 & 1.75 & 0.375 & 16.81  \\
   92023-03-08-10  & 54008.62108655  & 1.392 & 0.772 & 0.236 &   0 &   0 & 6.85 & 2.11 & 0.875 & 19.45  \\
   92023-03-10-10  & 54012.41419998  & 1.561 & 0.793 & 0.286 &   0 &   1 & 6.39 & 2.71 & 0.750 & 24.97  \\
   92023-03-16-10  & 54024.20448354  & 1.374 & 0.839 & 0.116 &   1 &   1 & 7.17 & 1.95 & 0.375 & 14.29  \\
   92023-03-20-10  & 54032.06032845  & 1.362 & 0.767 & 0.154 &   0 &   0 & 7.42 & 2.12 & 0.500 & 4.97  \\
   92023-03-27-10  & 54046.46863863  & 1.406 & 0.846 & 0.195 &   1 &   0 & 7.09 & 1.91 & 0.625 & 8.62  \\
   92023-03-34-10  & 54120.25891063  & 1.501 & 0.970 & 0.072 &   1 &   0 & 5.93 & 1.55 & 0.875 & 17.72  \\
   92023-03-44-10  & 54140.74365600  & 1.546 & 0.987 & 0.111 &   1 &   0 & 6.44 & 1.45 & 0.750 & 21.67  \\
   92023-03-44-00  & 54166.21890484  & 1.610 & 1.048 & 0.093 &   1 &   0 & 5.91 & 1.24 & 1.250 & 25.71  \\
   92023-03-64-00  & 54212.89935623  & 1.520 & 0.732 & 0.208 &   0 &   1 & 6.59 & 2.52 & 0.500 & -14.72  \\
   92023-03-64-10  & 54214.99848817  & 1.329 & 0.755 & 0.119 &   0 &   1 & 6.60 & 2.13 & 0.750 & 21.76  \\
   92023-03-70-00  & 54226.78303678  & 1.602 & 1.004 & 0.103 &   1 &   1 & 5.82 & 1.34 & 1.000 & 29.37  \\
   92023-03-71-00  & 54228.07044998  & 1.550 & 0.975 & 0.119 &   1 &   0 & 6.19 & 1.47 & 0.750 & 20.82  \\
   92023-03-73-00  & 54230.48778215  & 1.597 & 0.999 & 0.126 &   1 &   0 & 6.34 & 1.37 & 0.500 & 18.05  \\
   92023-03-65-10  & 54232.71368493  & 1.649 & 1.038 & 0.127 &   0 &   0 & 6.04 & 1.22 & 1.000 & 23.96  \\
   92023-03-66-10  & 54234.89705877  & 1.662 & 1.050 & 0.129 &   1 &   0 & 6.00 & 1.18 & 0.875 & 16.59  \\
   92023-03-73-10  & 54248.43857498  & 1.367 & 0.765 & 0.172 &   0 &   1 & 6.22 & 2.12 & 0.500 & -15.38  \\
   92023-03-76-10  & 54254.84918840  & 1.507 & 0.733 & 0.216 &   0 &   1 & 7.39 & 2.50 & 1.375 & 42.73  \\
   92023-03-83-10  & 54268.40871965  & 1.500 & 0.884 & 0.131 &   1 &   0 & 6.14 & 1.75 & 1.250 & 27.70  \\
  95337-01-02-000  & 55473.92615600  & 1.425 & 0.715 & 0.087 &   1 &   1 & 7.57 & 2.29 & 0.500 & 28.74  \\
  95337-01-02-000  & 55474.17558308  & 1.425 & 0.715 & 0.087 &   0 &   1 & 6.75 & 2.29 & 0.250 & 19.04  \\
  95337-01-03-000  & 55474.92021850  & 1.403 & 0.801 & 0.067 &   0 &   0 & 7.02 & 2.03 & 0.375 & 7.67  \\
   96322-01-05-02  & 55840.95671155  & 1.586 & 1.022 & 0.092 &   1 &   0 & 5.93 & 1.33 & 0.750 & 22.59  \\
  96322-01-05-000  & 55841.14026479  & 1.615 & 1.033 & 0.093 &   1 &   0 & 5.86 & 1.27 & 1.000 & 31.31  \\
  96322-01-05-000  & 55841.30065252  & 1.615 & 1.033 & 0.093 &   1 &   0 & 5.49 & 1.27 & 1.000 & 30.74  \\
   96322-01-05-00  & 55841.47716873  & 1.633 & 1.045 & 0.093 &   1 &   0 & 5.84 & 1.22 & 0.875 & 17.19  \\

\hline 
\end{tabular}
\label{tab:pileup_2}
\begin{tablenotes}
\item[]
\end{tablenotes}
\end{center}
\end{table*}

\section*{Acknowledgements}
This research has made use of data obtained from the High Energy
Astrophysics Science Archive Research Center (HEASARC), provided 
by NASA’s Goddard Space Flight Center. We thank Duncan Galloway
and  Yuri Cavecchi for useful comments and discussions. 
G.Z., M.Z. and A.C. are members of an International Team in Space 
Science on thermonuclear bursts sponsored by the International 
Space Science Institute in Bern, Switzerland. A.C. is supported 
by an NSERC Discovery Grant.



\end{document}